%% file: main.tex
\PassOptionsToPackage{dvipsnames}{xcolor}
\documentclass[trackchanges,twocolumn,resetfootnote]{aastex701}
\usepackage{tikz}
\usetikzlibrary{decorations.pathreplacing}

\usepackage{siunitx} 
\usepackage{journal_macros}
\usepackage{fontawesome5} 
\usepackage{amsmath}

\newcommand\externalgithublink[2]{\href{#2}{online \faGithub}}

\newcommand{\atantwo}[2]{\text{atan2}\left(#1, ~ #2\right)}

\newcommand{\nuon}{\nu_{t\_\text{on}}}
\newcommand{\nuoff}{\nu_{t\_\text{off}}}
\newcommand{\nuonoff}{\nu_{t\_\text{on/off}}}
\newcommand{\Eon}{E_{t\_\text{on}}}
\newcommand{\Eoff}{E_{t\_\text{off}}}

\newcommand{\Mon}{M_{t\_\text{on}}}
\newcommand{\Moff}{M_{t\_\text{off}}}
\newcommand{\Monoff}{M_{t\_\text{on/off}}}

\newcommand{\apepecc}{0.82}
\newcommand{\apepeccunc}{\pm0.04}
\newcommand{\apepphase}{0.35}
\newcommand{\apepphaseunc}{\pm0.02}
\newcommand{\apepoa}{126}
\newcommand{\apepoaunc}{\pm5}
\newcommand{\apepperiod}{193}
\newcommand{\apepperiodunc}{\pm11}
\newcommand{\apepinc}{24}
\newcommand{\apepincunc}{\pm3}
\newcommand{\apepexpansion}{1020}
\newcommand{\threeshells}{600}

\shorttitle{Dust Destruction in Apep}
\shortauthors{White et al}

\submitjournal{ApJ}

\begin{document}

\title[Dust Destruction in Apep]{The Serpent Eating Its Own Tail: Dust Destruction in the Apep Colliding-Wind Nebula}

\author[orcid=0009-0006-7054-0880,sname='White']{Ryan~M.~T.~White}
\affiliation{School of Mathematical and Physical Sciences, Macquarie University, Sydney, 2113, NSW, Australia}
\affiliation{School of Mathematics and Physics, University of Queensland, Brisbane, 4072, QLD, Australia}
\email[show]{ryan.white@mq.edu.au}  

\author[orcid=0000-0003-2595-9114,sname='Pope']{Benjamin~J.~S.~Pope} 
\affiliation{School of Mathematical and Physical Sciences, Macquarie University, Sydney, 2113, NSW, Australia}
\affiliation{School of Mathematics and Physics, University of Queensland, Brisbane, 4072, QLD, Australia}
\email{benjamin.pope@mq.edu.au}

\author[orcid=0000-0001-7026-6291, sname='Tuthill']{Peter~G.~Tuthill}
\affiliation{Sydney Institute for Astronomy, School of Physics, University of Sydney, Sydney, 2006, NSW, Australia}
\email{peter.tuthill@sydney.edu.au}

\author[orcid=0000-0002-2106-0403, sname='Han']{Yinuo~Han}
\affiliation{Division of Geological and Planetary Sciences, California Institute of Technology, 1200 E. California Blvd., Pasadena, CA 91125, USA}
\email{yinuo@caltech.edu}

\author[0000-0001-9145-8444,sname='Dholakia']{Shashank Dholakia}
\affiliation{School of Mathematics and Physics, University of Queensland, Brisbane, 4072, QLD, Australia}
\email{s.dholakia@uq.edu.au}

\author[0000-0003-0778-0321, sname='Lau']{Ryan~M.~Lau}
\affiliation{NSF NOIRLab, 950 N. Cherry Ave. Tucson, AZ 85719, USA}
\email{ryan.lau@noirlab.edu}

\author[0000-0002-7167-1819, sname='Callingham']{Joseph~R.~Callingham}
\affiliation{ASTRON, Netherlands Institute for Radio Astronomy, Oude Hoogeveensedijk 4, Dwingeloo, 7991 PD, The Netherlands}
\affiliation{Anton Pannekoek Institute for Astronomy, University of Amsterdam, Science Park 904, 1098\,XH, Amsterdam, The Netherlands}
\email{jcal@strw.leidenuniv.nl}

\author[0000-0002-2806-9339, sname='Richardson']{Noel~D.~Richardson}
\affiliation{Department of Physics and Astronomy, Embry-Riddle Aeronautical University, 3700 Willow Creek Road, Prescott, AZ 86301, USA}
\email{RICHAN11@erau.edu}

\begin{abstract}
    Much of the carbonaceous dust observed in the early universe may originate from colliding wind binaries (CWBs) hosting hot, luminous Wolf-Rayet (WR) stars. Downstream of the shock between the stellar winds there exists a suitable environment for dust grain formation, and the orbital motion of the stars wraps this dust into richly structured spiral geometries. The Apep system is the most extreme WR-CWB in our Milky Way: two WR stars produce a complex spiral dust nebula, whose slow expansion has been linked to a gamma-ray burst progenitor. It has been unclear whether the O-type supergiant 0.7" distant from the WR+WR binary is physically associated with the system, and whether it affects the dusty nebula. Multi-epoch VLT/VISIR and JWST/MIRI observations show that this northern companion star routinely carves a cavity in the dust nebula -- the first time such an effect has been observed in a CWB -- which unambiguously associates the O~star as a bound component to the Apep system. These observations are used together with a new geometric model to infer the cavity geometry and the orbit of the WR+WR binary, yielding the first strong constraints on wind and orbital parameters. We confirm an orbital period of over 190~years for the inner binary -- nearly an order of magnitude longer than the next longest period dust-producing WR-CWB. This, together with the confirmed classification as a hierarchical triple, cements Apep as a singular astrophysical laboratory for studying colliding winds and the terminal life stages of the most massive star systems.
\end{abstract}


\section{Introduction} 

Colliding wind binaries (CWBs) are some of the most extreme astrophysical environments known, where hot shocks are formed from the collision of massive stellar winds. A rare subclass of these systems harbour a Wolf-Rayet (WR) star together with a massive, main sequence OB companion. A handful of such binaries in the Galaxy are known to host WC subtype WRs and are often the site of copious dust production \citep{White2024arXiv, Lau2020ApJa}. This dust is produced from the shock of the colliding winds from the two stars, which offers an environment that nurtures dust nucleation through mixing of wind material at the contact discontinuity and self-shielding from the radiation of the massive stars \citep{Soulain2023MNRAS, Usov1991MNRAS}, although the exact mechanisms by which dust formation is mediated by the colliding winds remains an open question. The orbital motion of the two stars then wraps the outflowing dust plume into intricate spiral nebulae which encode the dynamical history of the stars, as well as their wind physics \citep{Tuthill1999Natur, Tuthill2008ApJ}. Although rare in the present-day Galaxy, these binaries are thought to have been a dominant contributor of carbonaceous dust in the Milky Way's early history and they offer a convincing explanation for structured circumstellar material seen in extragalactic supernova light curves \citep{Moore2023ApJ}. 

Of all the CWBs in the Galaxy, the Apep system stands out as the most puzzling; it is, at present, the only spectroscopically confirmed classical WR+WR system to host a WC type star, while its orbital period is the longest (by an order of magnitude) among the dusty CWBs \citep{Callingham2019NatAs, Callingham2020MNRAS}. In addition to, or perhaps because of this, the Apep system is a testbed of CWB physics across the electromagnetic spectrum; besides the infrared plume, the colliding winds result in unusually luminous non-thermal emission from high energies \citep[in the X-ray;][]{del-Palacio2023A&A, Zhekov2025A&A} to radio wavelengths \citep{Callingham2019NatAs, Bloot2022MNRAS, Driessen2024PASA}, with the non-thermal colliding wind shock itself having been imaged with very long baseline interferometry \citep{Marcote2021MNRAS} cementing it as the most luminous non-thermal CWB. These multiwavelength studies have propelled Apep into being one of the most well-studied CWBs even within the first 10 years of its discovery. 

One of the most intriguing open questions about the Apep nebula is the apparent discrepancy between the proper-motion expansion rate and the spectroscopic windspeeds of its component stars. Spectroscopy clocks the two winds at speeds of $2100\pm200$\,km\,s$^{-1}$ and $3500\pm100$\,km\,s$^{-1}$ (independent of a distance estimate), while the nebula expansion speed as measured by proper motions of the dust is much slower at $910\pm120$\,km\,s$^{-1}$ assuming a distance of 2.4\,kpc \citep{Callingham2020MNRAS, Han2020MNRAS}. Potential resolutions for these apparently contradictory values have been proposed, including an incorrect distance measurement, anisotropic winds in one or both of the central WR stars, complex wind braking physics at the shock front, or some combination of these. 

There is also some uncertainty as to whether the `northern companion' to Apep, an OIaf supergiant 0.7$^{\prime \prime}$ distant from the central WR+WR binary, is dynamically associated with the system. It is highly unlikely for a field star of this rare type to so closely coincide along the line of sight, and there is a similar dust extinction of this star compared to the central binary which suggests the same distance \citep{Callingham2020MNRAS}. Despite this, there has not yet been a definitive association of the star to the Apep system in large part due to the failure of previous studies to point to evidence of impacts on the dusty nebula as would be expected to be generated by a massive luminous star in close proximity.

Geometric modelling has so far been the workhorse tool in understanding the binary orbit of the ``Pinwheel'' CWBs \citep{Tuthill2008ApJ, White2024arXiv}, and indeed Apep's orbit. The goal of such modelling is to understand the geometric structure of the dust nebula and the orbit of the stars at its centre, while hydrodynamic modelling of CWBs is a more useful tool to understand the processes leading to dust formation \citep[e.g. as in][]{Eatson2022MNRASb, Soulain2023MNRAS}. The dust produced along the surface of the wind-wind shock is wrapped into a spiral from the orbital motion of the stars which can be modelled with a geometric prescription based on the orbital elements. 

Until now, only the single innermost dust shell of Apep has been observed (with the Very Large Telescope VISIR instrument), leading to ambiguity between orbital and nebula expansion parameters, some of which being highly covariant in model fits. Observations of older, concentric outer shells are capable of resolving this degeneracy. A dramatic illustration of this came recently from \emph{James Webb Space Telescope} (JWST) Mid-infrared Instrument (MIRI) instrument images that revealed over a dozen concentric shells of material around the WR\,140 system \citep{Lau2022NatAs, Lieb2025ApJ}. Given that each shell corresponds to one periastron passage of the central stars, this image represents $\sim 170$ years of dynamical history and yields an unambiguous orbital solution that agrees with the three-dimensional orbit of the system of \citep{Lau2022NatAs,Thomas2021MNRAS}. In the time since, four more WR CWBs have been observed with JWST: WR\,48a, WR\,112, WR\,125, and WR\,137 \citep{Richardson2025ApJ}, which has revealed the faint thermal emission of centuries old dust.

In this paper we present a strongly constrained orbital solution to the central WR+WR binary of the Apep system using the 3 nested concentric dust shells revealed by JWST MIRI imaging \citep{Han2025inprep}. We also employ these new data, together with multiple earlier epochs of VISIR imaging, to definitively link the O supergiant as a dynamically bound third member of the Apep hierarchical triple system through the cavity it carves in the wider nebula. These findings were generated by way of a state-of-the-art geometric code, described in Appendix~\ref{app:geom_model}, capable of modelling the impact of higher order effects such as dust acceleration and wind anisotropy.

\section{Methods}\label{sec:methods}
\subsection{Apep Direct Images}\label{sec:data}
The Apep system was observed with JWST MIRI as part of the JWST GO Program 5842 on July 24, 2024 for 3.64 hours, using the F770W, F1500W, and F2550W filters. We show a false-colour composite image using these filters in Figure~\ref{fig:apep_pretty} following the reduction process of \citet{Han2025inprep} and Section~\ref{sec:methods-miri-image}.

In addition to the multi-wavelength JWST images taken, Apep was imaged with the VLT VISIR instrument in June 2024, bringing the total number of VISIR-Apep epochs to 4 spanning the years 2016 to 2024. All VISIR images used in this study were taken in the J8.9 band which is centred at 8.7\,$\mu$m. The data reduction for the VLT images is described in more detail in \citet{Han2025inprep}.

\subsection{Geometric Model} \label{sec:geom-model}
We have developed a new and fast colliding wind nebula geometric model, \texttt{xenomorph} \citep{White2025Zenodo} -- which is motivated by the new VLT and JWST imagery of Apep -- that can be applied to any CWB with resolved dust spirals. The basic principle is similar to previous geometric codes, e.g. that in \citet{Han2022Natur}, although our new code is $\sim 100$ times more performant due to our using the \textsc{Jax} framework for Python \citep{jax2018github}, as well as featuring more physical parameters, being more user friendly, and better documented. The open-source code is available at \href{https://github.com/ryanwhite1/xenomorph}{github.com/ryanwhite1/xenomorph}. We describe the physical features of the code and their motivation in Appendix~\ref{app:geom_model}, and in this section we describe how we have used the software in the context of our Apep imagery.

\subsubsection{Parameter Estimation with the Geometric Model}
Geometric models are a key tool in learning about the orbital parameters of the binary stars at the heart of colliding wind nebulae. Fitting for these orbital and wind parameters involves forwards modelling the nebula, changing parameters until the ridge geometry best aligns with that which is observed; because of this, we need not consider the precise dust properties \citep[as described in the companion paper,][]{Han2025inprep}, only its density structure. That is, the simulated images reproduce the projected (and normalised) column density of dust in the nebula, and not its luminosity, temperature, or other quantities which require deeper modelling. 

The \texttt{xenomorph} code was designed with Bayesian parameter inference in mind, such as Markov Chain Monte Carlo (MCMC) or Hamiltonian Monte Carlo (HMC). However, there are too many parameters to fit for in using a conventional MCMC algorithm \citep[which struggle or fail with high dimensionality;][]{Huijser2015arXiv}, which leaves gradient-based methods like HMC as the only option. Although the code computes gradients in the model via autodifferentiation with \textsc{Jax}, the gradients are unstable and are not suitable for such a method. In addition to this, it is difficult to describe an informative goodness of fit function for the fitting, given that the geometric model does not account for the radiative transfer and thermal emission of the nebula (where doing so would make the computational cost intractable for fitting). With this in mind, we manually fit for the parameters using the difference imaging within the graphical user interface (GUI) built into the \texttt{xenomorph} model to best align the nebula structure. We describe below the parameter-specific choices we made in this fitting process.

\textbf{Orbital and system parameters.} We begin fitting for the imprint of the WR+WR binary orbit on the nebula using the previous best values in the literature \citep[most notably those in][]{Han2020MNRAS}. We treat the orbital eccentricity, inclination, longitude of ascending node, argument of periastron, and orbital phase as free parameters. We adopt the orbital period of \citet{Han2025inprep} which was found via a proper motion analysis of dust ridge expansion. The orbital semi-major axis and stellar masses are unconstrained by geometric modelling of the nebula at this spatial scale and so cannot be fit for. We adopt a distance of 2.4\,kpc \citep{Callingham2019NatAs, Han2020MNRAS} to the system which sets the nebula expansion speed as a fixed parameter when coupled with the adopted orbital period (due to the trigonometry of the nebula expanding in the plane of the sky). 

\textbf{Wind and dust parameters.} We treat the shock cone open angle (describing the momentum ratio of the two WR stellar winds), true anomaly of dust turn on/off, and azimuthal dust variation as free parameters in fitting. We fixed the amplitude of azimuthal variation to $A_\text{az}=0.5$ which appeared to best match the data. The orbital modulation and gradual turn on/off in dust production was not included in the fit in this study after initial tests with it did not appear to match the observations. 

\textbf{Tertiary cavity.} As described in Section~\ref{sec:results}, we have found strong evidence of a cavity in the north-eastern region of the Apep nebula. In constraining the geometry of this cavity (and hence the parameters of the tertiary star), we allowed the cavity open angle and the polar and azimuthal positions of the cavity to be free. The destruction constant was fixed to $A_\text{tert}=1.75$ although more negative values can just as suitably fit the observations also. In modelling the position of the stellar sprites, we also allowed the distance of the tertiary star from the inner binary to be a free parameter. Our proposed phenomenology for the cavity is that it is a result of sculpting from the tertiary companion in Apep, however in fitting for the position of the cavity we only considered the nebular geometry itself and not the position of the tertiary star in the imagery. 

\textbf{Miscellaneous parameters.} The nucleation distance of the dust with respect to the shock front is only constrained by high angular resolution observations of the inner core of colliding wind binaries. Given that our observations are over very large spatial scales, there is no way to constrain the nucleation distance from such direct images; regardless, the nucleation distance is only relevant in the geometric model at small spatial scales (on the order of the semi-major axis) as it dictates the distance at which dust first appears. Hence, we have fixed the nucleation distance to 0\,au and note that this has no effect on the results in an episodic dust producer and at an orbital phase far from periastron passage. Additionally, the brightness and width of the Gaussian profiles of the stars in the image were treated as free parameters, and were scaled until they appeared to roughly match the relative brightnesses in the imagery. Finally, all simulated images (including the difference fits of Figure~\ref{fig:apep-fit}) in this study are convolved with a Gaussian blur with a width of 2 pixels to emulate the observation point spread functions and the true `fluffiness' of the nebula edges.

\section{Results}\label{sec:results}
\begin{figure*}
    \centering
    
    \includegraphics[width=.9\textwidth]{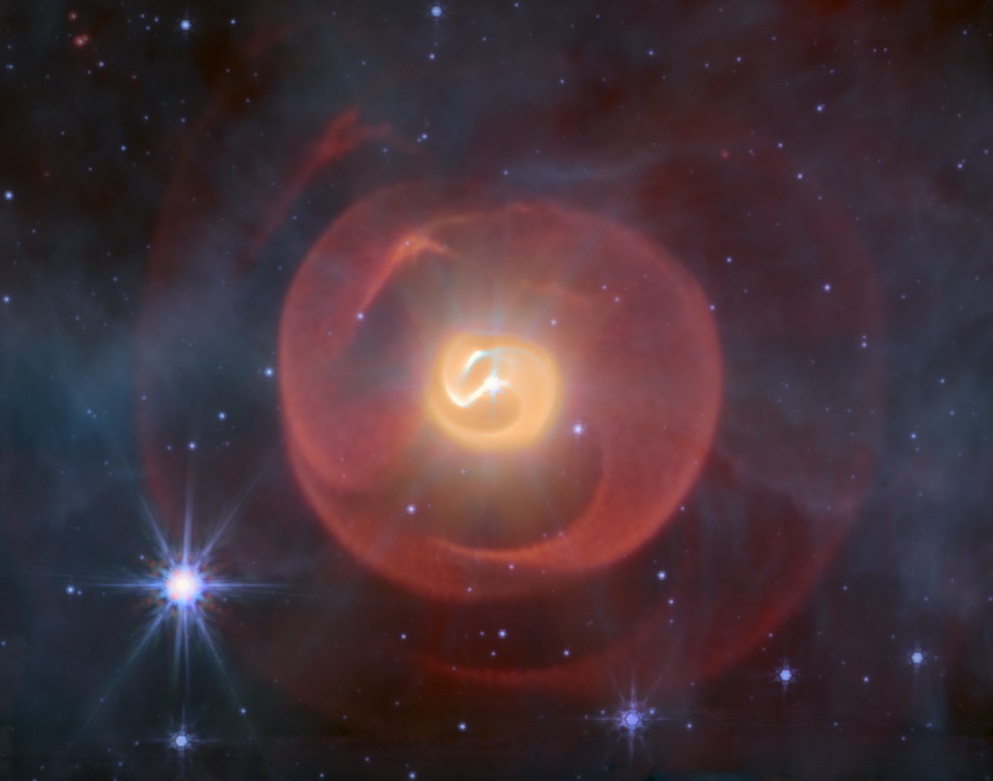}
    \makebox[\textwidth][c]{
    \hspace{1cm}
    \raisebox{1.07cm}{
    \includegraphics[width=.457\textwidth]{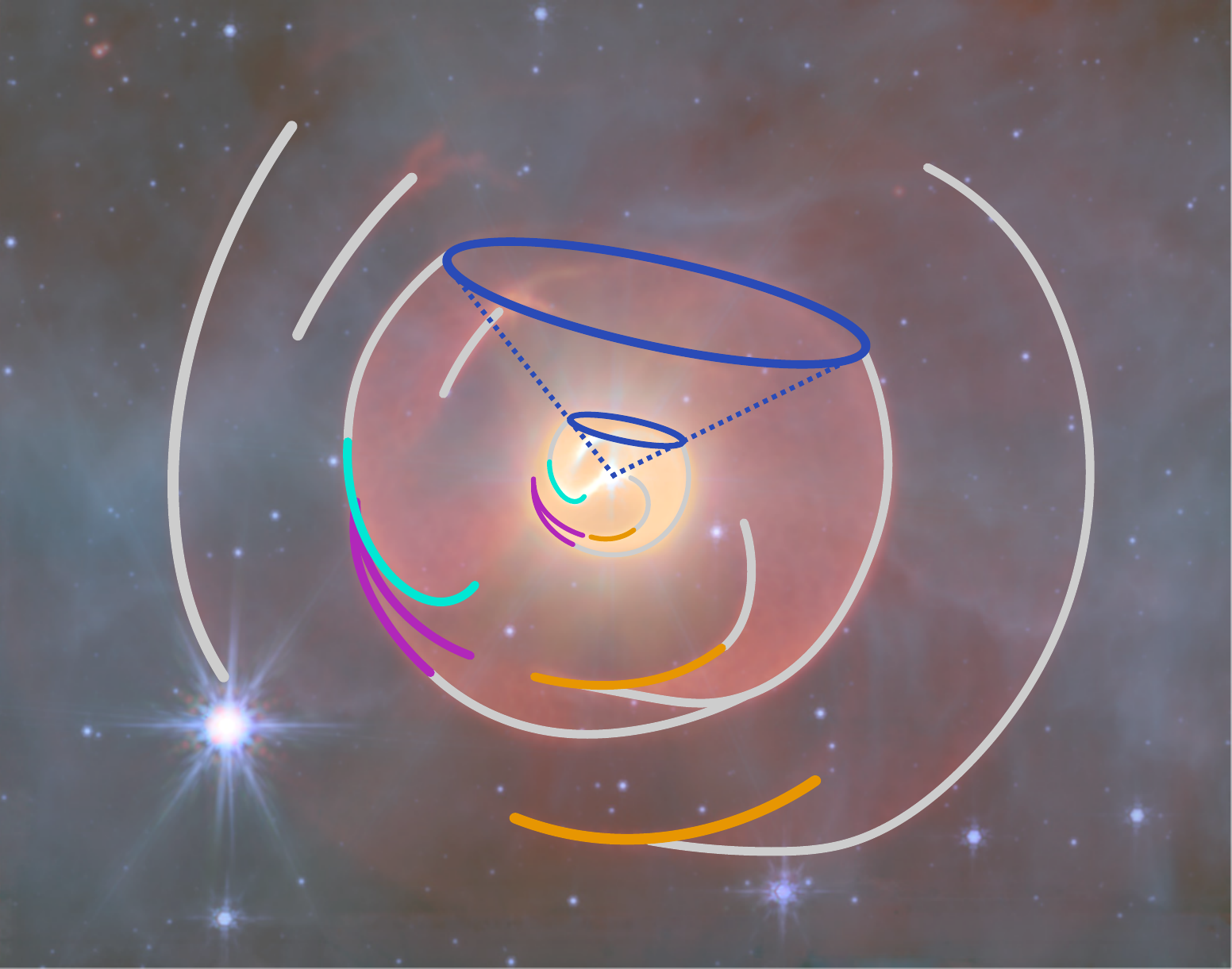}}
    \includegraphics[width=0.515\textwidth]{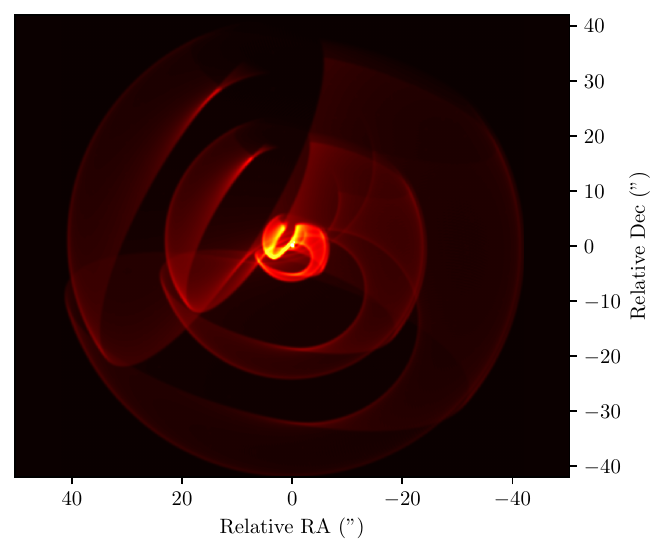}}

    \caption{Panel a) shows the false colour composite image of the Apep nebula made by combining data from the F770W, F1500W, and F2550W filters. Three concentric shells of dust are clearly seen, with faint evidence of a fourth at the edge of frame. The WR+WR central engine and the tertiary O supergiant lie in the centre of frame, while the bright star in the lower left is not associated with the Apep system. The processing procedure is described in Section~\ref{sec:methods-miri-image}. In panel b) we highlight the concentric shell structure with a grey skeleton, where the coloured ridges indicate regions of interest: the position of the \textcolor{Blue}{cavity} across concentric shells (see Figure~\ref{fig:apep-cone}), the relative positions of the \textcolor{TealBlue}{south-east ridge} coinciding with the \textcolor{magenta}{south-east tail} over time are key for the new eccentricity value, and the \textcolor{BurntOrange}{southern bar} aids in the period determination. Panel c) shows our geometric fit to the 3 visible shells of Apep, where the alignment of the modelled to the observed shells is shown in Figure~\ref{fig:apep-fit}.}

    \vspace*{-24cm}
        \begin{center}
            \hspace{-15cm}\textcolor{white}{a)}
        \end{center}
    \vspace*{24cm}
    
    \vspace*{-12cm}
        \begin{center}
            \hspace{-6.7cm}\textcolor{white}{b)}
            \hspace{8.2cm}\textcolor{white}{c)}
        \end{center}
    \vspace*{12cm}

    \vspace*{-18cm}
        \begin{center}
            \hspace{-12cm}
            \begin{tikzpicture}
                \input{compass_and_scale.tikz}
            \end{tikzpicture}
        \end{center}
    \vspace*{18cm}

    \label{fig:apep_pretty}
\end{figure*}

\begin{figure*}
    \centering
    \includegraphics[width=\linewidth]{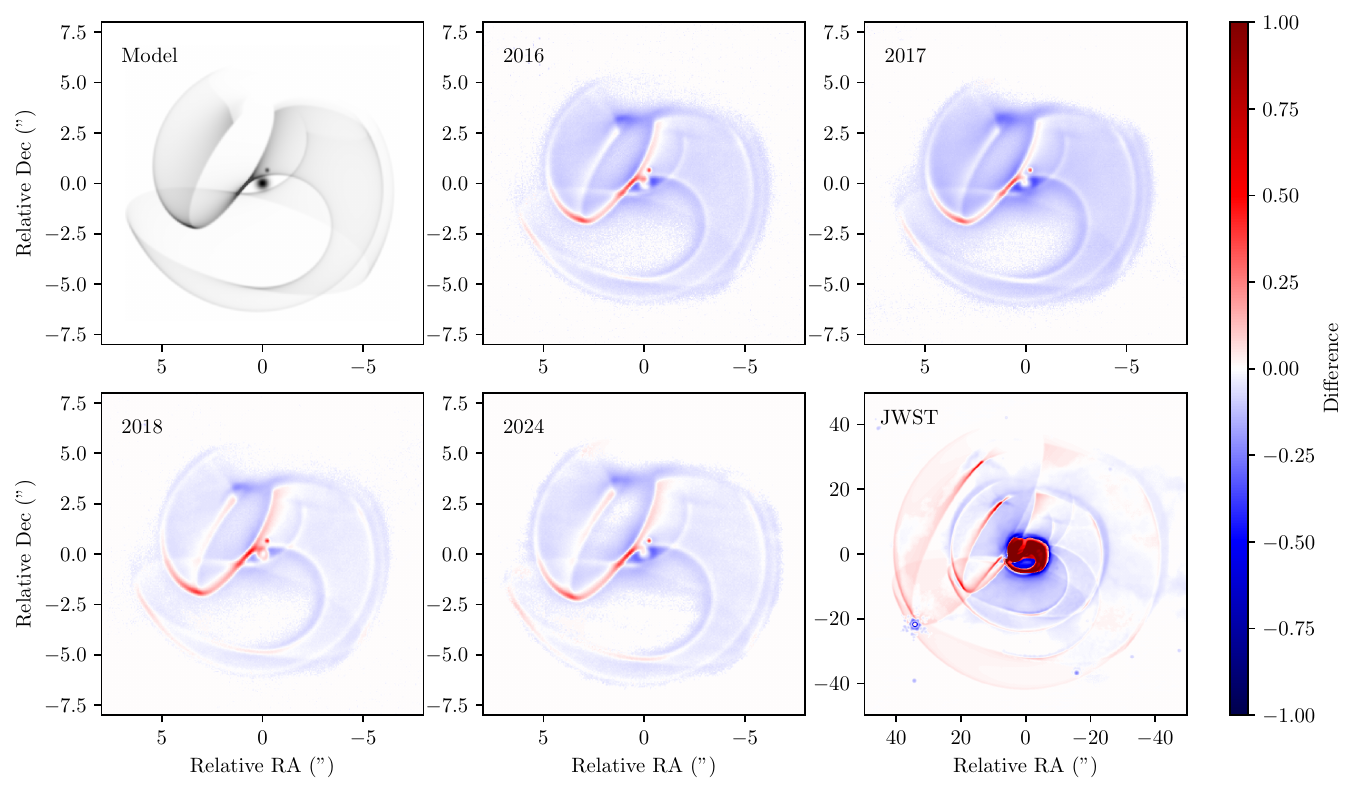}
    \caption{Using the 4 VLT epochs plus the 25.5\,$\mu$m JWST image, we found parameters for our geometric model that faithfully reproduce the observed geometry; the numerical values of each parameter are shown in Table~\ref{tab:apep-params}. The top left panel shows the model geometry at $\phi_{2024}$, and all other panels show the difference imaging of $F_\text{model} - F_\text{obs}$ adjusted for the respective orbital phases; note that the JWST panel (bottom right) is over a larger angular scale. Although the exact flux often does not entirely cancel out in the difference imaging, we note the nebula ridge positions -- which encode all of the fit parameters -- align very well. We have an animation available \externalgithublink{Apep evolution over orbital period}{https://github.com/ryanwhite1/ryanwhite1.github.io/blob/5d6c4c84b9a65402912bed784c18a0c8b09488b6/permanent\_storage/apep\_photos/Apep\_evolution.gif} that shows the evolution of a shell with our parameters over one orbital period.}
    \label{fig:apep-fit}
\end{figure*}

In this paper we are primarily concerned with modelling the geometric structure of the Apep nebula and how this constrains the binary orbit and stellar parameters. Figure~\ref{fig:apep_pretty} reveals three concentric shells of dust that closely match expanded versions of a single common structure, although with evidence for some degree of geometric evolution between them. The innermost shell cannot be mapped by a pure inflation to perfectly match the the older (outer) shells since the projected geometry evolves slightly over time as the three dimensional structure expands (see the linked animation in the Figure~\ref{fig:apep-fit} caption). Each shell is the result of dust production coinciding with a periastron passage, so this image contains nearly a millennium of dynamical information about the Apep system.

To analyse the structure of the Apep nebula, we developed a new computationally efficient geometric model that reproduces the dust morphology given orbital and stellar parameters (see Section~\ref{sec:geom-model}). By fitting simulated structures to the images, we determined the system parameters shown in Table~\ref{tab:apep-params} with the fits to the images shown in Figure~\ref{fig:apep-fit}. The fitting was aided by multiple epochs of Apep imagery spanning 8~years from the VLT, and the VLT images together with the JWST image allowed us to break any degeneracies between orbital phase, orbital period, and nebular expansion speed. When fitting with only a single epoch and a single shell, the ridge geometry can be suitably fitted by a range of degenerate orbital periods, expansion speeds and phases; more epochs of imaging and more visible shells thus greatly constrains the parameter space. The most surprising result of the fitting \citep[and the proper motion analysis in][]{Han2025inprep}, given by the separation between successive shells now revealed by JWST, is that the central WR+WR binary in the centre of the Apep system must have an orbital period of $\apepperiod\apepperiodunc$\,yr -- almost twice as long as previously thought based on analysis of the single innermost shell. We note that this period estimate is agnostic of distance -- as it only relies on the shell spacing and angular expansion speed -- although the geometric fitting of the period is covariant with expansion speed; simultaneously fitting the VISIR epochs + the MIRI image places a lower bound of 170\,yr for the orbital period, where periods below this cannot match the observed geometry evolution. This makes Apep by far the longest period colliding wind binary to produce a dusty nebula at almost an order of magnitude longer than WR\,48a \citep[whose period is $\sim32$\,yr;][]{Williams2012MNRAS}. The shell spacing requires that the current orbital phase of the inner WR+WR binary be $\phi_{2024} = 0.35\pm0.02$ at the time of the JWST observation; this means that the last periastron passage would have occurred in the year $1956\pm6$ and the next in $2149\pm9$. Interestingly, the date of previous periastron passage is consistent with the previous estimate using the period of 125\,yr and phase of $\sim0.6$ given in \citet{Han2020MNRAS}. The conditions that allow copious dust production -- and indeed its extended survival -- in such a widely separated long period system are not well understood, but the unique configuration of a double WR star at the heart of Apep is likely a pivotal factor.   

\begin{table*}
    \centering
    \renewcommand{\arraystretch}{1.2}
    \begin{tabular}{ccc}
        \hline
        Parameter & Value & Reference \\
        \hline
        Eccentricity, $e$ & $\apepecc\apepeccunc$ & W25 \\
        Inclination, $i$ & $(\apepinc\apepincunc)^\circ$ & W25 \\
        Long. of Asc. Node, $\Omega$ & $(164 \pm 15)^\circ$ & W25 \\
        Arg. of Periastron, $\omega$ & $(10\pm 10)^\circ$ & W25 \\
        Opening Angle, $\theta_\text{OA}$ & $(\apepoa\apepoaunc)^\circ$ & H20, W25 \\
        Orbital Period, $P_\text{orb}$ & $\apepperiod\apepperiodunc$\,yr & H25, W25 \\
        Orbital Phase at July 2024, $\phi_{2024}$ & $\apepphase\apepphaseunc$ & W25 \\
        Distance, $d$ & $2.4^{+0.2}_{-0.5}$\,kpc & C19, H20 \\
        Dust Shell Expansion Speed, $v_\text{dust}$ & $\apepexpansion \pm 100$\,km\,s$^{-1}$ & H25, W25 \\
        Dust Turn On True Anom., $\nuon$ & $(-108\pm10)^\circ$ & W25 \\
        Dust Turn Off True Anom., $\nuoff$ & $(141\pm 10)^\circ$ & W25 \\
        Azimuthal Variation, $\sigma_\text{az}$ & $(30 \pm 10)^\circ$ & W25 \\
        \hline
    \end{tabular}
    \caption{Our best-fitting parameters for the Apep system. The opening angle parameter $\theta_\text{OA}$ is slightly updated from the reference, and we make no attempt to fit for the distance to the system. The dust expansion speed, $v_\text{dust}$, is consistent with the 90\,mas\,yr$^{-1}$ found by the proper motion analysis in \citet{Han2025inprep} when at a distance of 2.4\,kpc. References: C19 -- \citet{Callingham2019NatAs}; H20 -- \citet{Han2020MNRAS}; H25 -- \citet{Han2025inprep}; W25 -- This work.}
    \label{tab:apep-params}
\end{table*}

With successive outer dust shells now visible thanks to the sensitivity of JWST MIRI, we are able to better understand how the dust geometry changes over time. Comparing the innermost (first) shell to the second shell, we can see that the relative spacing between the south-east ridge and south-east tail is reduced over time (Figure~\ref{fig:apep_pretty}, panel b). This, together with the significantly earlier orbital phase estimate at 2024 $\phi_{2024}$, means that the WR+WR binary must be on a more eccentric orbit to have reduced that separation by such a degree over \apepperiod~years. We have found that $e=0.82\pm0.04$ most accurately reproduces this ridge positioning across concentric shells. This propels Apep into a class of high-eccentricity CWBs such as WR\,140 or WR\,19 \citep{Williams2009MNRAS, Williams2009MNRASb}, where the conditions at the shock should change significantly over the course of the orbit. Still, the physical processes at the colliding wind shock in Apep are different from that of other WR CWBs; Apep's shock should be adiabatically cooled throughout its whole orbit while systems such as WR\,140 are radiatively cooled, particularly at periastron \citep{Parkin2008MNRAS}. This has implications as to the conditions of dust production throughout Apep's orbit. With our parameters we find that dust production `turns on' at $\phi \simeq 0.97$ and `turns off' at $\phi \simeq 0.10$, which translates into a production period of $\sim 25$\,yr. We do not find any evidence of orbital modulation in dust production -- as seen in WR\,140 \citep{Han2022Natur} -- but some azimuthal variation in dust production (i.e. the trailing edge of the shock with respect to orbital motion favours dust production) makes for a closer geometric fit to the observed shells, consistent with WR\,140 \citep{Han2022Natur, Lau2022NatAs}.

\begin{figure}
    \centering
    \includegraphics[width=\linewidth]{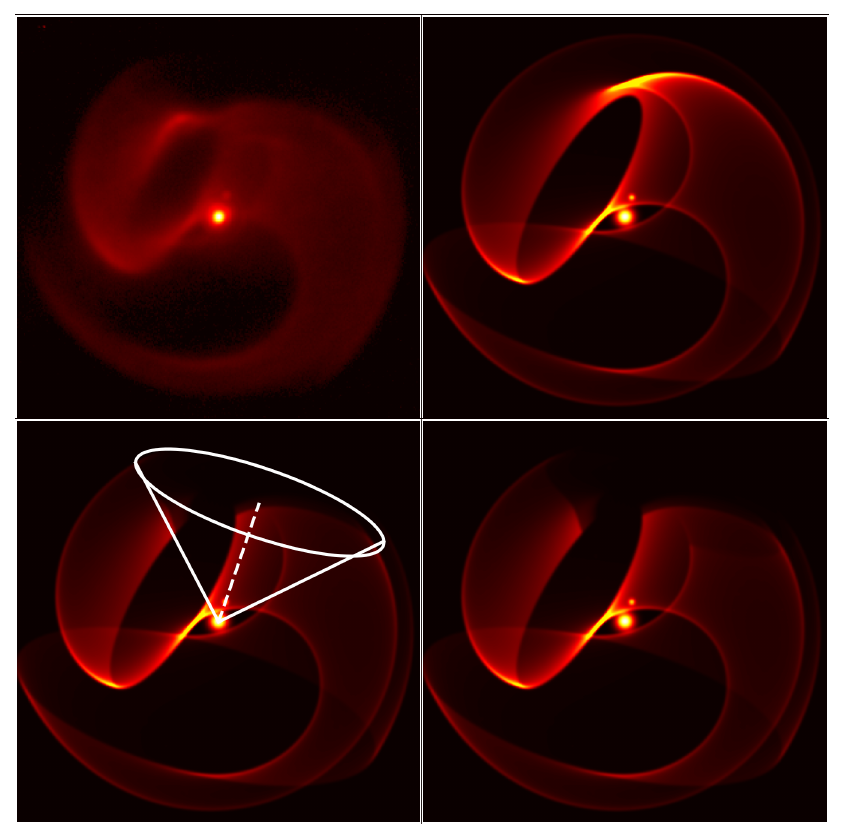}
    \caption{An illustration of the cavity carved by the northern O star companion in the Apep system. The WR+WR binary is the central bright point, while the northern companion is the smaller point source 0.7" to the north. \emph{Top left:} The innermost shell of the Apep nebula observed with VLT/VISIR, processed as in Section~\ref{sec:methods-visir-image}. \emph{Top right:} Previous models with no dust destruction effects from the northern companion overestimate the dust luminosity in the upper region of the plume. \emph{Bottom left:} We model dust destruction along the volume of a cone extending out of the WR+WR central engine along the line-of-sight vector to the northern companion. \emph{Bottom right:} With the addition of this region of dust destruction, the model geometry now more closely matches what we see with direct imaging (top left).}
    \label{fig:apep-cone}
\end{figure}

\begin{table}
    \centering
    \renewcommand{\arraystretch}{1.2}
    \begin{tabular}{ccc}
        \hline
        Parameter & Value \\
        \hline
        Inclination Angle, $\beta_\text{tert}$ & $(124\pm10)^\circ$ \\
        Azimuthal Angle, $\alpha_\text{tert}$ & $(239\pm 10)^\circ$ \\
        Opening Angle, $\theta_\text{OA,tert}$ & $(90\pm 10)^\circ$ \\
        Tertiary Star Radial Position, $r_\text{tert}$ & $1700 \pm 200$\,au\\
        \hline
    \end{tabular}
    \caption{Parameters describing the simple conical model for the dust depletion region of Apep due to the northern companion in an ordinate system relative to the WR+WR binary orbital plane. An inclination angle of $0^\circ$ lies directly below the orbital plane, while the periastron position of the inner binary is at $180^\circ$ azimuthal angle. For the $r_\text{tert}$ parameter, we assume the distance to Apep is $d = 2.4$\,kpc.}
    \label{tab:apep-photodissociation}
\end{table}

\subsection{Physics of the Dust Cavity} \label{sec:cavity}
One of the key open questions since the discovery of Apep is whether or not the `northern companion' star, an O8\,Iaf supergiant 0.7" north of the central WR+WR binary, is associated with the WR stars and Apep nebula, or is a (statistically unlikely) apparent line-of-sight alignment. In analysing archival VLT data, together with the new VLT epochs and JWST imagery, we have conclusively linked the O star to the WR stars, identifying the Apep system as a hierarchical triple. The new evidence for this lies in an apparent `cavity' of dust along the direction of the tertiary star from the WR+WR binary. Fig~\ref{fig:apep-cone} shows that by incorporating a `dust destruction' region along the surface of a cone into our geometric model, we may accurately explain the missing dust. The parameters of the destruction cone, in spherical coordinates relative to the orbital plane of the WR+WR binary, are shown in Table~\ref{tab:apep-photodissociation}. We also note that adding a stellar sprite 1700\,au from the inner binary at the 3D angular position of the cavity precisely matches the observed location of the tertiary star in Apep, further supporting the O star as the cavity sculptor. Due to the asymmetric nature of the Apep nebula, the angular position of the O star is uniquely associated with the position of the cavity and so there are no alternative configurations that can explain such cavity ridge positions. 

Our proposed mechanism for this cavity is as follows. As the dust forms from the central WR+WR colliding wind shock, that material expanding radially northwards enters the environment of the O star companion. A large fraction of this northbound dust is destroyed and no longer contributes, carving a gap in the shell geometry. The dust then continues outward where the cavity persists as a scar indicative of earlier interaction with the tertiary star. Rather than destroyed, we investigated several scenarios where the dust is displaced but none could reconcile the observed geometry in the nebula (this is discussed further in Section~\ref{sec:dust-grains}). 

The root cause of the apparent cavity of dust in the Apep nebula is almost certainly a combination of several processes in parallel. From our derived geometry of the cavity, we are able to discern some relative dominance between competing effects. The main pieces of information we have pertaining to the dust physics is that the cavity open angle is approximately $90\pm10^\circ$, and that the O supergiant companion lies $1700\pm200$\,au from the WR+WR central engine -- with these two pieces of information together, we find that the radius of the cavity when the dust shell directly passes the O companion is also about 1700\,au. 

The first avenue to dust destruction we consider is grain sublimation as a result of radiative heating from the O star. For a UV source of brightness $L_{\text{UV}}$ and grain sublimation temperature of $T_{\text{sub}}$, \citet{Hoang2019NatAs} gives the sublimation distance as 
\begin{equation}
    r_{\text{sub}} \simeq 0.015 \left(\frac{L_{\text{UV}}}{10^9 L_{\odot}}\right)^{1/2} \left(\frac{T_{\text{sub}}}{1800\,\text{K}}\right)^{-2.8}\,\text{pc} \label{eq:subdistance}
\end{equation}
For carbonaceous dust with sublimation temperature $\sim 1800$\,K \citep{Lau2023ApJ}, and a source of UV luminosity $\sim 10^5\,L_{\odot}$, we arrive at a sublimation radius of $\sim 30$\,au. This is two orders of magnitude smaller than the required dust destruction radius of $\lesssim 1700$\,au, and so grain sublimation cannot be the dominant process of dust destruction in the Apep nebula. 

The second process of destruction we consider is by grain-grain or grain-ion collisions inducing grain shattering. O supergiant stars, like Wolf-Rayets, are characterised by fast and dense winds, so we would expect a significant wind-wind shock where the nebula intercepts the supergiant wind. To test whether we observe such a shock as expected, we consider the wind momentum ratio, 
\begin{equation}
    \eta = \frac{\dot{M}_1 v_{\infty,1}}{\dot{M}_2 v_{\infty,2}} \label{eq:momentumratio}
\end{equation}
where $\dot{M}$ and $v_\infty$ are the mass loss rates and terminal windspeed of each star, and how this momentum ratio relates to the shock opening angle
\begin{equation}
    \eta_{\text{S/W}} = \left(\frac{121}{\theta_{\text{OA}}/2} - \frac{31}{90}\right)^3 \label{eq:momentumratio-openangle}
\end{equation}
given by \citet{Tuthill2008ApJ, Gayley2009ApJ}, where $\eta_{\text{S/W}}$ is the ratio of the stronger to the weaker wind, and $\theta_{\text{OA}}$ is the shock opening angle. Inserting an opening angle of $90^\circ$ yields a momentum ratio of $\eta_{\text{S/W}}\sim 13$ for the mixed WR+WR wind to the O8\,Iaf wind. If a wind-wind shock is the dominant influence on this cavity, we would expect to recover a similar wind momentum ratio with equation~\ref{eq:momentumratio}. For this, we can use our nebular expansion speed of \apepexpansion\,km\,s$^{-1}$ and estimate a mass flux of $\sim 10^{-4}-10^{-4.75}\,M_{\odot}$\,yr$^{-1}$; the exact mass flux is difficult to discern since it is the mixed product of two WR winds, each with different intrinsic mass loss rates, but each should be independently in this range \citep{Callingham2020MNRAS}. \citet{Callingham2020MNRAS} gives the terminal windspeed of the O star as $1280\pm50$\,km\,s$^{-1}$, and if we assume a mass loss rate in the range $10^{-5.2}-10^{-6}\,M_\odot$\,yr$^{-1}$ \citep[typical for O supergiants of this spectral type;][]{Crowther2009A&A}, we can recover a wind momentum ratio of $\eta_{\text{S/W}}\sim 13$ using equation~\ref{eq:momentumratio}. This result implies that the cavity geometry may be explained at least partially by the presence of a tertiary shock with grain-ion sputtering. Although the radius of the cavity is 1700\,au, the tertiary shock front will be significantly closer to the O star at the point of stagnation between the mixed WR+WR wind and the O star wind. Following \citet{Marcote2021MNRAS, Canto1996ApJ}, the stagnation point will be at a distance
\begin{equation}
    R_\text{W/S} = \frac{r_\text{tert}}{1+\eta_\text{S/W}^{1/2}}
\end{equation}
from the O star (with the weaker wind). Inputting our distance and wind momentum ratio values gives a distance of $\sim 370$\,au, which is still an order of magnitude too large for the dust destruction mechanism to be from dust sublimation. In Section~\ref{sec:dust-grains} we describe more mechanisms that may be working in parallel on the dust grains within the cavity geometry.

\section{Discussion and Conclusions}

In light of the finding of accelerating dust shells of WR\,140 \citep{Han2022Natur}, we incorporated dust acceleration for the first time into a geometric model. We chose a phenomenological prescription that can account for a range of smooth acceleration profiles as described in Appendix~\ref{sec:dust-accel}. We find that there is no evidence of dust acceleration nor deceleration in the dusty shells of the Apep nebula. The former result is expected, since this acceleration is expected to be due to radiation pressure significant only at very small orbital phases \citep[which are not probed in our imagery;][]{Han2022Natur}, while the latter result indicates there are negligible drag forces acting on the shells as they expand into the ISM. This in turn implies that the strong winds from this WR system kinematically dominate the surrounding region at least to the limits of the shells we see. Our imagery also indicates that the dust from the Apep nebula is capable of surviving for at least $3P_\text{orb} \simeq \threeshells$\,yr while subjected to the harsh ISM environment.

Since the discovery paper of the Apep nebula, there has been a puzzling contradiction between the measured nebular expansion and the spectroscopic wind speeds of the central WR stars \citep{Callingham2019NatAs}. Our fitted expansion speed of $\apepexpansion\pm100$\,km\,s$^{-1}$ \citep[at an assumed distance of $2.4^{+0.2}_{-0.5}$\,kpc;][]{Han2020MNRAS, Han2025inprep} confirms this discrepancy. A proposed resolution to this is that at least one of the WR stars in the system has a slow and dense equatorial wind with a fast polar wind, such that the observed slower uniform expansion of the shells and fast winds may coexist. To investigate this we incorporated a phenomenological anisotropic wind into the geometric model that alters the expansion speed and opening angle of some sections of the simulated shell so as to emulate orbital phase-dependent wind conditions as different latitudes of the WR stellar surface contribute to the wind-wind shock (Appendix~\ref{sec:anisotropy}). Such a model cannot rule out or confirm anisotropic winds where the stellar rotation is aligned with orbital motion. However, using this we have found that it is unlikely for the (would-be) anisotropic star to have its rotation significantly misaligned with the orbital plane as we do not find any geometric signatures of this in the nebula. If the slow-wind region extends to high stellar latitudes, we would also not be able to detect such a shell deformation. This is consistent with previous spectroscopic observations of a fast polar wind, given the plane of the orbit is almost in the plane of the sky (inclination of $\apepinc\apepincunc^\circ$). 
Investigating this further with hydrodynamic simulations will be essential to determine if Apep really does harbour a critically rotating star that may be a long gamma-ray burst (LGRB) progenitor.

\subsection{Effect on Dust Grains in the Cavity}\label{sec:dust-grains}

Our tertiary cone modelling does not exactly produce the observed structure and one of the last remaining mysteries in the Apep plume geometry is the origin of the horizontal ridge outlining the cavity on its southern border. Our first interpretation of this ridge is that the O star may deflect some proportion of the dust onto this outer ridge, but modelling this has proven difficult. We considered and implemented several models of deflection in an attempt to explain this ridge, where the O star deflects dust particles: into a ring along the entire cavity edge; onto the cavity edge on those angular coordinates co-located with the dust; or onto a central angular position along the cavity boundary with some azimuthal spread, but all three models were unsuccessful in reproducing the geometry. The first two ideas significantly over-predict ridges at other locations that are not seen in the observations, and the third can produce a ridge in the correct location (by design) but at an angle which does not match the observed ridge. Given that this is a persistent feature across the shells, we do not think that this is an anomalous dust feature (created by e.g. a surface mass eruption). Consequently, our picture of the tertiary shock is that it does not displace dust grains but rather destroys them. Estimates on the dust grain properties within and around the cavity are difficult to discern without spectra, however, but constraints can be inferred based on the environmental conditions.

The disruption of dust grains via radiative torques (radiative torque disruption -- RATD) has recently been studied in the context of Wolf-Rayet colliding wind binaries \citep{Lau2023ApJ, Hoang2019NatAs}. This occurs due to high energy photon absorption inducing rotation in asymmetric dust grains, thereby inducing a centrifugal stress which can disrupt the grains \citep{Hoang2019NatAs}. At a given distance $r_*$\,pc from a central point source of luminosity $L_*$ (units of $10^9 L_\odot$) with mean wavelength $\bar{\lambda}_{0.5}$ (in units of $0.5\,\mu$m), the maximum size of dust grains with tensile strength $S_9$ (units of $10^9$\,erg\,cm$^{-3}$) surviving this mechanism is given by \citet{Hoang2019NatAs} as
\begin{equation}
    a_{\text{RATD}} \simeq 0.003197 \left(\bar{\lambda}_{0.5}^{-1.7} L_*^{-1/3} r_*^{2/3} S_9^{1/2}\right)^{1/2.7} ~ \mu\text{m}
\end{equation}
Using this, together with our cavity edge distance of 1700\,au, a mean wavelength given by a blackbody distribution of 30\,kK, a luminosity of $10^{5.9}\,L_\odot$ (typical of O8\,Iaf supergiants), and a grain maximum tensile strength of $10^9$\,erg\,cm$^{-3}$ \cite[for disordered grains,][]{Hoang2019ApJ}, we obtain a maximum surviving dust grain size of $\sim 4$\,nm. The timescale on which this destruction takes place is 
\begin{equation}
    \tau_{\text{RATD}} \simeq \frac{368}{438} \, \bar{\lambda}_{0.5}^{-1.7} L_*^{-1} r_*^2 S_9^{1/2} \left(10\, a_\text{RATD}\right)^{-0.7} \, \text{yr}
\end{equation}
which, with the same parameters and calculated maximum grain size, is $\sim 6$\,yr \citep{Hoang2019NatAs}. More ordered, stronger grains such as graphite can have tensile strengths of $\sim 2 \times 10^{10}$\,erg\,cm$^{-3}$ \citep{Hoang2019ApJ}, however, which results in a maximum grain size of $\sim 13$\,nm and destruction timescale of $\gtrsim 12$\,yr. We note that the maximum grain sizes and destruction timescales due to the RATD mechanism could be underestimated by factors of a few given our orders of magnitude uncertainty in the tensile strengths across the true grain distribution. We also note that these calculations are done given parameters at the periphery of the dust cavity; in reality, we might expect a gradient of dust conditions across the cavity `surface'. These calculations assume that the environmental conditions are constant across the destruction timescale, when this is not the case: the dust shell, produced from the colliding wind of the WR+WR binary, is moving radially outwards first into the O stars wind and radiation and then out of it. In parallel the dust is becoming less dense as it expands into a larger shell surface with time, implying that the dust grains will be less shielded from radiation as time goes on (which explains in part why the tertiary star apparently destroys so much dust compared to the similar radiation environment of the central WR+WR binary). 

Our overall picture of the dust destruction resulting in a circular cavity consists of several processes in parallel. We suggest that the tertiary wind-wind shock breaks down large grains into nano-grains via grain-grain and gas-grain collisions, while the radiation pressure from the O supergiant disrupts large grains above $\sim 4$\,nm into nano-grains via the RATD mechanism. The closest $\lesssim 1700$\,au to the O tertiary will see an ionising environment up to orders of magnitude more intense than the photon-dominated regions around sites of massive stellar formation (mainly due to the much smaller length scales involved in the cavity, albeit on shorter timescales), and so nano-grains will be photodissociated by the UV flux as they are formed \citep{Schirmer2022A&A}. The many competing phenomena warrant the use of detailed simulations, for example with a hydrodynamical code, to model this observed geometry in Apep in future studies. 

\subsection{Orbit of the Tertiary Star}
If we take the distance to the Apep system to be $2.4$\,kpc \citep{Callingham2019NatAs, Han2020MNRAS}, the tertiary star must be presently $1700\pm200$\,au from the WR+WR binary centre of mass. Given the positioning of the dust cavity, the O star is approximately $34\pm10^\circ$ above the plane of the WR+WR orbit. If the tertiary star is in a circular orbit at such a distance, with a total system mass of $\sim 80M_\odot$, the orbital period should be of order $P_\text{tert} \sim 8 \times 10^3$\,yr. If this is the case, we would expect the tertiary star -- and by extension its associated cavity in the dust plume -- to move $\sim 28^\circ$ in its orbit around the common CoM over the 3 visible dust shells and $3P_\text{orb}\simeq \threeshells$\,yr of visible dust in the JWST imagery. This extent of cavity displacement is not supported by our fits which is constrained to $\lesssim 10^\circ$ of movement. Imaging of even older shells -- which would be cooler, older, and larger -- with the Atacama Large Millimetre Array (ALMA) may unambiguously reveal the movement of this cavity over a longer timescale. Similarly, detection of molecular lines, and their subsequent radial velocity profile, in the Apep nebula with ALMA would yield an unambiguous expansion velocity (and corresponding orbital period) of the nebula. 

The slow movement of the cavity can be explained if the O star companion is on an eccentric orbit, where it is presently near apastron and moving slower relative to an analogous circular orbit. This is a plausible explanation, since often in hierarchical triples the companion star is on a highly elliptical, distant orbit \citep{Toonen2016ComAC, Antognini2016MNRAS}. In such a case, the tertiary star will have a perturbative influence on the dynamics of the inner binary. Even with a modest eccentricity of $e_\text{tert} = 0.1$, the Kozai-Lidov timescale of this configuration would be of order $P_\text{KL}\sim 10^5$\,yr \citep{Naoz2016ARA&A, Antognini2015MNRAS}. Massive stars -- which eventually evolve into WR stars -- have lifetimes of a few million years, and so the Apep system will have experienced several Kozai-Lidov timescales at minimum; since this mechanism can in principle pump the eccentricity of the inner binary up to values of $e \gtrsim 0.95$, the companion star must be accounted for in any binary evolution studies of the Apep system in order to discern past epochs of close passages and/or mass transfer. 

\subsection{Concluding Remarks}
The creation and processing of dust grains into the interstellar medium influences all areas of astronomy, and the binary WR stars together with the dust-destroying O supergiant at the heart of the Apep nebula cements it as a truly unique astrophysical laboratory. The JWST observations of Apep reveal luminous circumstellar dust that is several times the age of that revealed around WR\,140 in previous observations \citep{Lau2022NatAs, Lieb2025ApJ}, and support our finding that the O supergiant `northern companion' is dynamically associated with the binary WR stars in Apep; this is the first time that dust destruction has been observed by a tertiary star in a colliding wind nebula, and marks Apep as part of a rare class of triple colliding wind binaries together with WR\,147 \citep{Rodriguez2020ApJ}. Our work motivates more detailed simulations of the dust destruction in such a system, for example with a hydrodynamic code, and the modelling of dynamical effects of a tertiary star on the binary evolution in Wolf-Rayet colliding wind binaries.  

\section{Data Availability}
The JWST data used in this study were obtained from the Mikulski Archive for Space Telescopes (MAST; \href{https://archive.stsci.edu/missions-and-data/jwst}{https://archive.stsci.edu/missions-and-data/jwst}) at the Space Telescope Science Institute, under JWST GO program ID 5842. The JWST data used in this paper can be found in MAST: \dataset[doi:10.17909/mf14-ga19]{http://dx.doi.org/10.17909/mf14-ga19}.

\section{Code Availability}
The geometric model code, \texttt{xenomorph} \citep{White2025Zenodo}, is open source on GitHub and available at \href{https://github.com/ryanwhite1/xenomorph}{github.com/ryanwhite1/xenomorph}, with a documentation site online at \href{https://ryanwhite1.github.io/xenomorph/}{ryanwhite1.github.io/xenomorph/}. We encourage community contributions to the code through pull requests, and any suggestions or feedback on the state of documentation. The code makes use of \textsc{Jax} \citep{jax2018github}, NumPy \citep{Harris2020Natur}, Matplotlib \citep{matplotlib}, SciPy \citep{SciPy2020NaturMth}, and Astropy \citep{Astropy2013A&A, Astropy2018AJ, Astropy2022ApJ}. The version of the code current to the date of publication is available on Zenodo and can be accessed via \dataset[doi:10.5281/zenodo.15875502]{https://doi.org/10.5281/zenodo.15875502}.

\begin{acknowledgments}
We are grateful to Peredur Williams, Orsola De Marco, and Katelyn Smith for valuable discussions during the analysis and writing of this work. We thank the University of Sydney and University of Queensland `FOCI' group for helpful discussions and feedback throughout the project. RMTW acknowledges the financial support of the Andy Thomas Space Foundation. SD and BJSP were funded by the Australian Government through the Australian Research Council DECRA fellowship DE210101639. NDR is grateful for support from the Cottrell Scholar Award \#CS-CSA-2023-143 sponsored by the Research Corporation for Science Advancement.

This work is based on observations made with the NASA/ESA/CSA \emph{James Webb Space Telescope}. The data were obtained from the Mikulski Archive for Space Telescopes at the Space Telescope Science Institute, which is operated by the Association of Universities for Research in Astronomy, Inc., under NASA contract NAS 5-03127 for JWST. These observations are associated with program \#5842. Support for program \#5842 was provided by NASA through a grant from the Space Telescope Science Institute, which is operated by the Association of Universities for Research in Astronomy, Inc., under NASA contract NAS 5-03127.

We acknowledge and pay respect to the traditional owners of the land on which the University of Queensland is situated, upon whose unceded, sovereign, ancestral lands we work. We acknowledge the traditional custodians of the Macquarie University land, the Wallumattagal clan of the Dharug nation, whose cultures and customs have nurtured and continue to nurture this land since the Dreamtime. We acknowledge the tradition of custodianship and law of the Country on which the University of Sydney campuses stand. We pay our respects to those who have cared and continue to care for Country.

\end{acknowledgments}

\begin{contribution}
RMTW developed the geometric wind model, conducted all data analysis, and led the writing of the manuscript. YH led the telescope proposals and contributed to discussions and analysis. BJSP and PGT supervised the project and provided comments on the manuscript. SD and BJSP reduced and processed MIRI imagery for both scientific and aesthetic purposes.
All authors contributed to the text of the final manuscript.

\end{contribution}

\facilities{JWST(MIRI), VLT(VISIR)}

\appendix

\section{Image Data Processing}
\subsection{For the Poster Image} \label{sec:methods-miri-image}
We use the software Pixinsight\footnote{\href{https://pixinsight.com/}{https://pixinsight.com/}} for the generation of the image shown in Figure~\ref{fig:apep_pretty}, aimed at generating a false-colour composite image for visual recognition of the shell structures and other features in the image. We start by cropping the image and filling in parts of the image missing data by interpolation with surrounding pixels. We then construct the false-color composite by assigning the channels red, green and blue respectively to F2550, $0.5\times\text{F770}+0.5\times\text{F1500}$, and F770. A rectangular segment containing little dust or stars is set as a background reference and the colours are transformed such that this is set to a neutral grey colour tone. We then apply a custom nonlinearisation to the image to reveal the background shells, and increase the colour saturation. Lastly, we use the Python package starnet++\footnote{\href{https://www.starnetastro.com/}{https://www.starnetastro.com/}}, a neural-network based tool to mask the stars and apply a custom colour transformation to get closer to the appearance of a blackbody. We emphasise that the aim of these processes is for aesthetic and visual purposes rather than scientific. 

\subsection{For Geometric Fitting} \label{sec:methods-visir-image}
The fitting of geometric forms to data relies primarily on the alignment of ridges present in both the model image and the observed data. Therefore, we minimally processed the raw VISIR and MIRI data to accentuate the ridge geometry. To do this we performed a 64th percentile background subtraction on each image then normalised all pixel values to be between 0 and 1 (by taking a floor of 0 brightness and then dividing by the maximum value). We then raised all pixel values to the 0.5th power which emphasised the nebula brightness suitably for fitting. The reference images in Figure~\ref{fig:apep-fit} (shown by the blue shading) have been processed in this way.

\section{The \texttt{xenomorph} Geometric Model} \label{app:geom_model}
\begin{figure}
    \centering
    \includegraphics[width=0.4\linewidth, trim=0 0.7cm 0 0, clip]{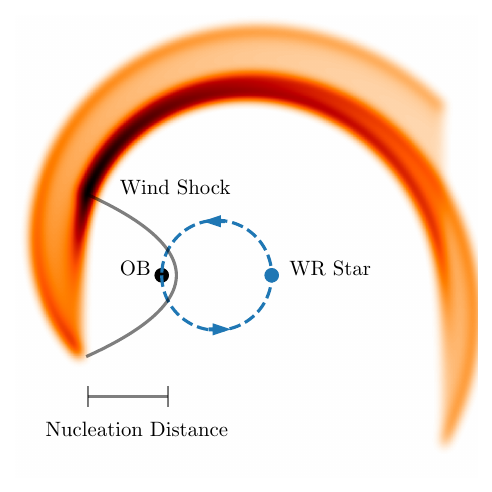}
    \includegraphics[width=0.4\linewidth, trim=0 0 0 2.8cm, clip]{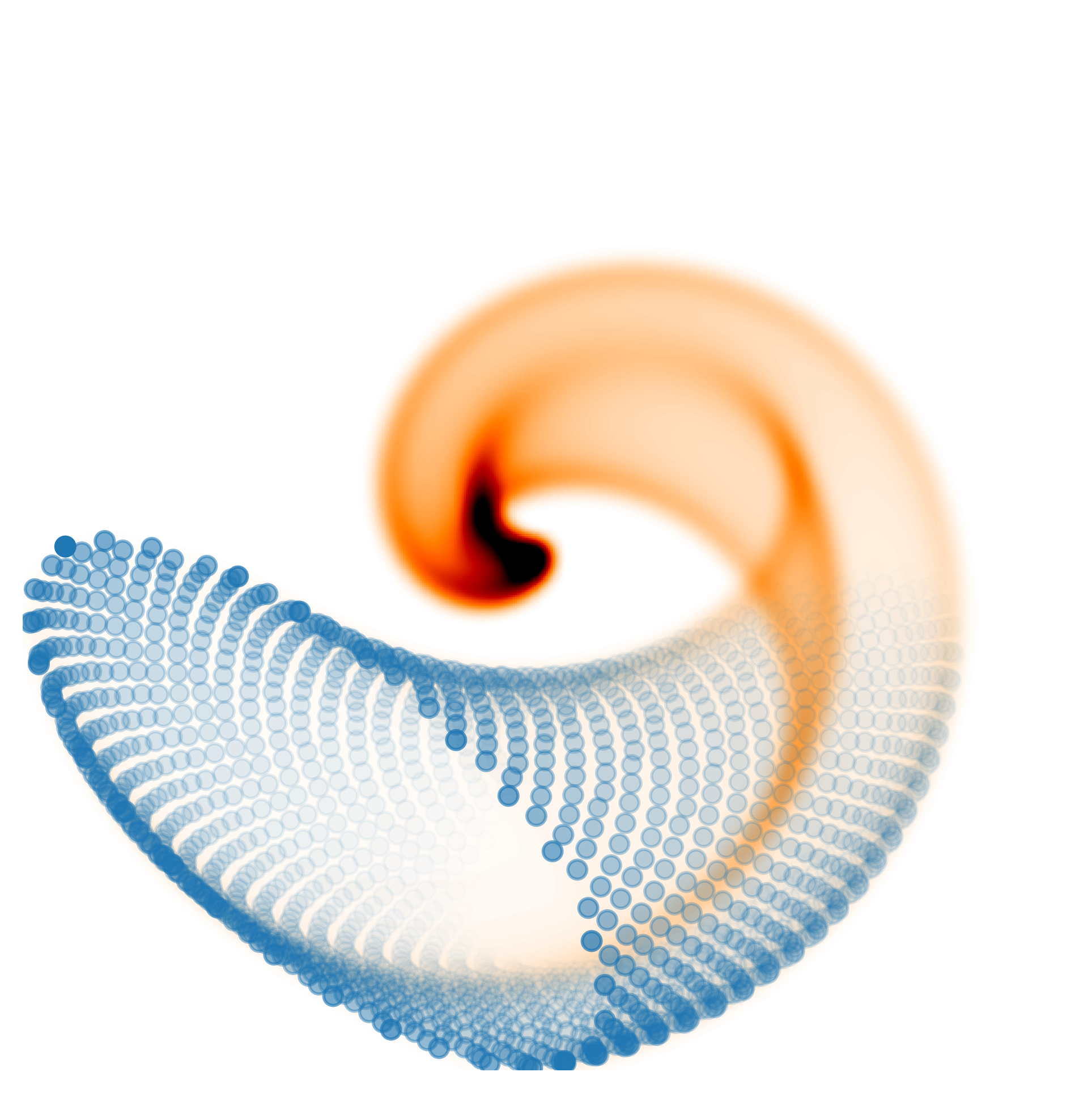}
    \caption{The idea for the geometric model is to initialise a ring of points around the wind-wind shock at some nucleation distance behind the secondary star (usually an OB-type star). As the binary orbits the common centre of mass, the formed dust (and discretised rings/particles in the right figure) gets wrapped into a spiral as it expands out from the central source. Note that there are orders of magnitude fewer particles (compared to the default code setting) in the right-side plot for visualisation purposes.}
    \label{fig:basic_idea}
\end{figure}
The basic idea that underpins the geometric model is to approximate instantaneous dust production within the wind collision region as a ring expanding from the stagnation point down the surface of the wind-wind shock cone. This ring is initialised in the orbital plane at the location of the `secondary' star (in terms of wind strength, not mass as is convention), and ejected along the line of sight from the primary star to the secondary at the wind speed of the primary. As the two stars orbit each other, the direction of ring ejection changes with the result that the plume `wraps' to create the characteristic spiral. This is illustrated in Figure~\ref{fig:basic_idea}. 

Even at this basic level, the model has a multitude of parameters that can be changed at run time. The parameters most fundamental to this geometric approximation are the number of discretised points within each ring, and the number of rings to produce over the course of one orbital period. Ideally both of these would be as large as possible in order to reduce the error associated with approximating a continuous process as discrete (so-called quantisation error); practically we usually set these to be 400 points per ring and 1000 rings per orbit which provides a dense point cloud for the nebula. 

To render the point cloud into an image that resembles the real nebulae, we take a histogram of the line of sight projection of the cloud. Even though these nebulae are three dimensional, in imaging we observe the column density of the dust due to its projection onto the plane of the sky. Hence by collapsing our modelled 3D point cloud to a 2D image we can reproduce the observed images. By utilising the weighting functionality of histograms, we can selectively alter the importance of selected points in order to emulate physical phenomena responsible for sculpting the nebula (e.g. dust turn off, azimuthal variation, etc.).

Previous geometric models have been successful in reproducing the structure of the spiral nebula, although they are computationally inefficient. If parameter inference is the goal of using these models, we need to make the model as computationally fast as possible so as to reduce compute time, both for interactive fitting and particularly to enable machine optimisation with an MCMC. Further, the high dimensionality of these problems requires that gradient-based, Hamiltonian Monte Carlo (HMC) methods be used so that the model can converge to a valid solution in finite time. Fortunately, the \textsc{Jax} Python framework \citep{jax2018github} offers automatic differentiation capability for Python code and comes standard with Just In Time (JIT) compilation that speeds up model evaluation significantly. Therefore, creating a geometric model from scratch with \textsc{Jax} allows us to satisfy all needed criteria at once. Indeed, our complete code takes of order $\lesssim 0.1$\,seconds to generate the hundreds of thousands of particles in the point cloud and render them into an image, compared to that of $\sim 5$\,s with an equivalent implementation in stock \texttt{NumPy} \citep{Harris2020Natur}.

\subsection{The Basic Ring Model} \label{sec:ring_model}
Before rings can be created, we first need a model of the binary orbit so that we know where to initialise the rings and in what direction they move. The motion of two celestial bodies in their mutual orbit is described by Kepler's equation \citep[for a review of the two-body problem]{Murray1999ssd}. To calculate the true anomaly of a star at any point over the orbital period (i.e. with known mean anomaly, $M$), Kepler's equation
\begin{equation}
	M = E - e \sin E \label{eq:kepler}
\end{equation}
must be solved for the eccentric anomaly, $E$. Traditionally, this is solved by an iterative method which converges to the correct value of $E$ given $M$. Those iterative methods are not differentiable, however, and so we use the non-iterative method described in \citet{Markley1995CeMDA} and implemented in \textsc{Jax} in \citet{jaxoplanet}.
From the eccentric anomaly, we can then calculate the true anomaly, $\nu$, analytically via
\begin{equation}
	\nu = 2 \, \atantwo{\sqrt{1 + e} \sin (E / 2)}{\sqrt{1 - e} \cos (E / 2)}
\end{equation}
Since the mean anomaly changes linearly with time, this calculation of true anomaly provides the angular position of each body relative to the focus of their orbits (i.e. the centre of mass) at constant time intervals.

The choice of mean anomaly grid is directly proportional to the time domain over which dust is produced. If the system in question produces dust constantly, a grid choice encompassing $0 \leq M \leq 2\pi$ then results in the relevant true anomalies with which to initialise rings. If the system is an episodic dust producer, i.e. produces dust between a well-defined dust `turn on' and `turn off', the choice of a mean anomaly grid is not so straightforward. Previous implementations of the geometric model would initialise rings over a full circle of mean anomaly but then discard rings whose true anomaly fall out of the specified turn on/off range. As producing `invisible' rings requires significant computational work -- the same compute time as the visible rings and to no effect on the visible nebula -- we avoid this and devise a method to produce only rings that contribute to the visible spiral plume.  

To produce rings only between dust turn on/off, we need to find a grid of mean anomalies between our true anomaly bounds; essentially the inverse problem to that described above. 
We begin with our true anomaly bounds, $\nuon$ and $\nuoff$. For true anomaly we follow the convention that periastron corresponds to $\nu = 0$ and so we expect that the turn on is in the range $-\pi \leq \nuon \leq 0$ radians, conversely $0 \leq \nuoff \leq \pi$ radians for turn off anomaly. The eccentric anomaly corresponding to dust turn on is then calculated as
\begin{equation}
    \Eon = 2\, \atantwo{\tan \left(\frac{\nuon}{2}\right)}{\sqrt{\frac{1 + e}{1 - e}}}
\end{equation}
and the same is used to calculate $\Eoff$ instead with $\nuoff$. The result of this can be substituted into Equation~\ref{eq:kepler} to then obtain the mean anomaly bounds corresponding to dust turn on/off, $\Monoff$. For the episodic dust producers, we create a grid of mean anomalies in the range $\Mon \leq M \leq \Moff$ for each single shell which is fed back into the solver to find the true anomalies corresponding to the stellar orbits. 

In order to eventually determine the size and distance of each ring from the binary centre, the age of each ring must be calculated. An equally spaced grid within the mean anomaly bounds naturally provides an equally spaced grid in time. The age of the ring corresponding to each mean anomaly value, relative to the orbital period (i.e. in the range $0 \leq t_\text{age} \leq 1$), can then be calculated by
\begin{equation}
    t_\text{age} = \text{mod}\, \left(\frac{\text{mod}\, \left(M, ~ 2\pi\right)}{2\pi} - \phi, ~ 1\right) \label{eq:ring_age}
\end{equation}
where $\phi$ is the current orbital phase of the binary. Equation~\ref{eq:ring_age} results in rings created at the current orbital phase having the lowest age, and those at the previous mean anomaly grid position having the largest age. When multiple shells of material are being simulated, the grids of mean anomalies and ages are tiled $N_\text{shell}$ times, where each successive tiling has an additional $+1$ to the age which accounts for the previous orbital periods of time. To convert from this nondimensionalised age relative to the orbital period into an absolute time, we can simply multiply by $P_\text{orb}$. 

Using the true anomalies corresponding to dust production, the positions of the stars in their orbits can be calculated so that the positions and orientation of each ring can be found. The nondimensionalised position of the primary star is calculated as
\begin{equation}
    \mathbf{\hat{r}}_1 = (\cos (\nu), ~ \sin (\nu), ~ 0) \label{eq:position}
\end{equation}
for each true anomaly value, where we note that we take the orbital motion to be in the $x-y$ plane; the corresponding position of the secondary is just the negative of Equation~\ref{eq:position}, $\mathbf{\hat{r}}_2 = - \mathbf{\hat{r}}_1$, since angular displacement of the stars is always $\pi$ radians relative to the system barycenter. To convert this into a true distance, we multiply by the distance of each star relative to the system barycenter, $\vec{r}_{1/2} =  r_{1/2} \,\mathbf{\hat{r}}_{1/2}$, where
\begin{equation}
    r_{1/2} = a_{1/2} (1 - e \cos (E))
\end{equation}
for each corresponding eccentric anomaly to the true anomaly, and for each star. From Kepler's Third Law, the semi-major axis of the binary system is 
\begin{equation}
     a = \left(\frac{G (M_1 + M_2) P_\text{orb}^2}{4\pi^2}\right)^{1/3}
\end{equation} 
and so the semi-major axis of each star in their orbits relative to the system barycenter is
\begin{equation}
    a_1 = a\, \frac{M_2}{M_1 + M_2}; \qquad a_2 = a\, \frac{M_1}{M_1 + M_2}
\end{equation}
Calculating the absolute positions of the stars, especially relative to each other, is necessary when determining the position and angle of each ring relative to the system. 

We construct the rings along the $x$-axis (so that a circle is seen in the $y-z$ plane), and populate them with $N_\text{particles}$ particles per ring; this means that there are a total of $N = N_\text{rings} \times N_\text{particles}$ particles in the `point cloud' of each nebula shell. The particles are linearly spaced within the ring and constitute a full circle, such that $0 \leq \theta_\text{particle} \leq 2\pi$. The coordinates of the particles within each ring depends on the plume opening angle, $\theta_\text{OA}$, by 
\begin{equation}
    \mathbf{\hat{r}}_\text{particle} = \left(\cos \left(\frac{\theta_\text{OA}}{2}\right), ~ \sin \left(\frac{\theta_\text{OA}}{2}\right) \sin (\theta_\text{particle}), ~ \sin \left(\frac{\theta_\text{OA}}{2}\right) \cos (\theta_\text{particle})\right)
    \label{eq:ring_particles}
\end{equation}
The particle coordinates within each ring are then scaled by the distance they should have travelled given the ring age and the WR dust expansion speed, 
\begin{equation}
    \vec{r}_\text{particle} = v_\text{dust} \, t_\text{age} \, \mathbf{\hat{r}}_\text{particle} \label{eq:ring_expand}
\end{equation}
This expands the ring along the surface of a cone projecting outwards from the system centre, and both enlarges the ring and displaces it by the appropriate amounts.  The direction that this cone is facing is calculated from the current stellar positions as 
\begin{equation}
    \varphi = \atantwo{\frac{y_2 - y_1}{||\vec{r}_1 - \vec{r}_2||}}{\frac{x_2 - x_1}{||\vec{r}_1 - \vec{r}_2||}} \label{eq:ring_direction}
\end{equation}
which is the angle from the primary to secondary star in the orbital plane. Since the rings are created along the $x$-axis, they must be rotated about the $z$-axis by $\varphi$ so that they align with the direction vector of the primary to the secondary. That is, we update each particle vector to be
\begin{equation}
    \vec{r}_\text{particle} = \left(R_z (\varphi) \vec{r}_\text{particle}^{\,T}\right)^T = \left(\begin{bmatrix}
        \cos (\varphi) & - \sin (\varphi) & 0 \\ \sin (\varphi) & \cos (\varphi) & 0 \\ 0 & 0 & 1   \end{bmatrix} \begin{bmatrix}
            x_\text{particle} \\ y_\text{particle} \\ z_\text{particle}
        \end{bmatrix}\right)^T
        \label{eq:ring_rotate}
\end{equation}
While Equations~\ref{eq:ring_particles}--\ref{eq:ring_rotate} are described in terms of a single particle, these operations are applied to all of the particles that belong to a single ring at once. This procedure is made computationally efficient by the \texttt{vmap} functionality within \textsc{Jax}, which automatically vectorises the particle and ring mathematics.

In our simulations we create each ring with $N_\text{particles} = 500$ particles and each shell with $N_\text{rings} = 1000$ rings. These numbers were chosen as a suitable middle ground between computational performance (which would favour fewer rings and particles) and dense spatial sampling of the plume (which would require as many rings and particles as possible). Since the true colliding wind nebulae are smooth to first order, our approximation of a point cloud introduces quantisation error into any rendered image. Hence, our choice of $N_\text{particles}$ and $N_\text{rings}$ was motivated so that there would always be a small but non-zero number of particles in even the lowest density rendered pixels.  

With our point cloud generated, next we need to render an image of the simulated nebula so as to compare it to observations. For this, we collapse our 3D point cloud into two dimensions which is analogous to its projection in the plane of the sky. We then take a 2D histogram of these collapsed particle locations which produces a `heatmap' of the nebula. Although this is a simple procedure, producing a column density of particles in this way is essentially what we observe when taking real astronomical images of optically thin material: the line of sight projection of material in the field of view. As a by-product, using a histogram allows us to very easily change the number of bins in the render as a proxy of changing the angular resolution of the simulation. This proves useful when comparing a simulated image to that taken by a telescope; we can render the image with the exact number of pixels (bins) needed on the correct angular pixel scale by manually fixing just two parameters.

Since the image is rendered in two dimensions, we must first ensure that the plume geometry is consistent with the orbital elements prior to rendering. That is, we must rotate the entire point cloud in 3D with the appropriate Euler angles so that the point cloud resembles the true geometry from our (observer's) perspective. We update the coordinates of each particle within the point cloud, $\vec{r}_\text{particle}$, as
\begin{align}
    \vec{r}_\text{particle} &= \left(R_z (-\Omega) R_x (-i) R_z (-\omega) \vec{r}_\text{particle}^{\,T}\right)^T
\end{align}
where $\Omega$ is the binary orbit longitude of ascending node, $i$ the inclination, and $\omega$ the argument of periastron. Note that these parameters do not affect the intrinsic geometry of the nebula, but only its projection into the plane of the sky.

\subsection{Azimuthal and Orbital Variation in Dust Production}
The first additional feature to the model that we consider is the azimuthal variation in dust production on each created ring. WR\,140 was observed to create dust most strongly on the trailing edge (with respect to the orbital motion) of its shock cone in \citet{Williams2009MNRAS}. In the time since, this phenomenon has been required to reproduce the observed geometry of WR\,140 in multi-epoch imaging including with JWST \citep{Williams2009MNRAS, Han2022Natur, Lau2022NatAs, Lieb2025ApJ}. This enhanced dust production along the trailing edge of the shock has also been supported by hydrodynamical simulations of WR\,104 \citep{Lamberts2012A&A, Soulain2023MNRAS}, and also more generally in the case of CWBs on eccentric orbits with non-equal winds \citep{Lemaster2007ApJ}.

Since we render each point cloud into an image via a histogram, we can apply weightings to each particle which affects how much it contributes to the final geometry. Utilising this functionality, we model this azimuthal variation as a Gaussian function multiplier applied to each particle within a ring and each entire ring respectively, similarly to \citet{Han2022Natur}. That is, the weighting factor of each particle from azimuthal variation is calculated by
\begin{equation}
	\delta w_\text{az} = \text{max} \left\{1 - (1 - A_\text{az}) ~ \exp \left(-\frac{(\theta_\text{particle} - \theta_\text{az\_min})^2}{2\sigma_\text{az}^2}\right), ~ 0 \right\} \label{eq:azimuthal}
\end{equation}
where $A_\text{az}$ is the global strength of the variation, $\theta_\text{particle}$ is the angular coordinate of the particle within the ring ($0^\circ$ being in the $+z$ direction), $\theta_\text{az\_min}$ is the location of the azimuthal minimum and $\sigma_\text{az}$ is the spread of the variation across the angular coordinates of the particles. Although each of these are free parameters, we usually set $\theta_\text{az\_min} = 90^\circ$ in practice; this corresponds to the leading edge of the dust plume with respect to the orbital motion resulting in dust production being weaker at the leading edge. A minimum value of 0 in Equation~\ref{eq:azimuthal} is enforced since $A_\text{az}$ can in principle take any positive value, although it is usually in the range $0 \lesssim A_\text{az} \leq 1$. This multiplicative factor calculated for all of the particles within the ring, and we approximate this effect as being constant across orbital phases. 

\begin{figure}
    \centering
    \includegraphics[width=.75\linewidth]{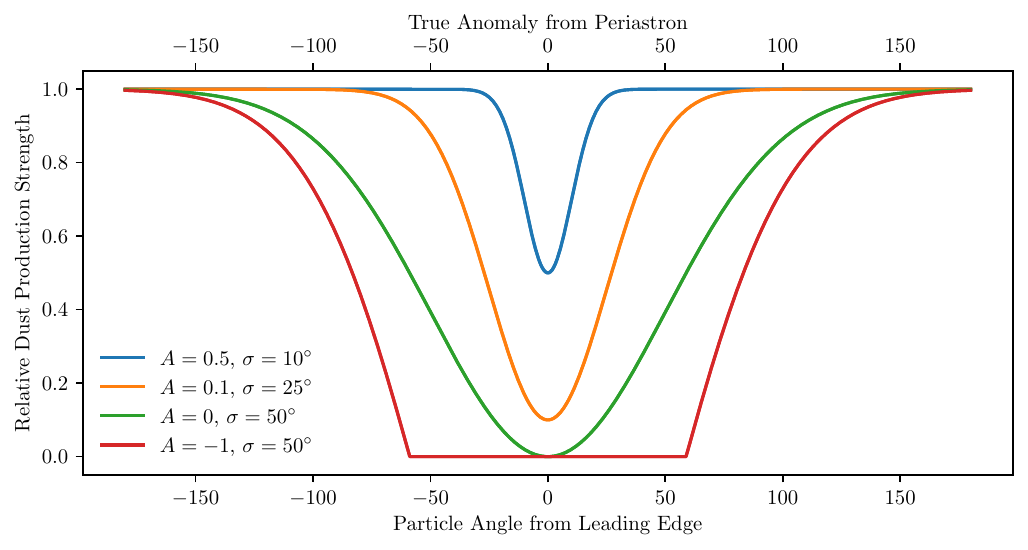}
    \caption{The free parameters of Equations~\ref{eq:azimuthal} and \ref{eq:orbital} change the depth and breadth of the strength variations in dust production. The top axis represents the weighting applied to each ring as a function of true anomaly for a representative set of free parameters, while the bottom axis represents the weighting applied to each particle within each single ring. }
    \label{fig:variation_gaussian}
\end{figure}

The dust nebula of WR\,140 shows evidence not only of azimuthal asymmetry, but orbital modulation in dust production between the classical turn on and off \citep{Williams2009MNRAS, Han2022Natur}. This manifests as a dip in the dust production when the binary stars are near periastron which results in two distinct phases of dust nucleation per orbit: at ingress and egress. We model this orbital modulation much in the same way as for the azimuthal variation,
\begin{equation}
	\delta w_\text{orb} = \text{max} \left\{1 - (1 - A_\text{orb}) ~ \exp \left(-\frac{(\nu_\text{ring} - \nu_\text{orb\_min})^2}{2\sigma_\text{orb}^2}\right), ~ 0 \right\} \label{eq:orbital}
\end{equation}
which is applied to an entire ring's weighting. The parameters in this Gaussian are analogous to that of the azimuthal variation, instead using the true anomaly of the entire ring instead of a particle angular coordinate, and with orbital variation amplitudes and spreads instead of azimuthal. Although technically free, we set $\nu_\text{orb\_min} = 180^\circ$, corresponding to the periastron of the orbit where the dust production may be weak during the dust production phase.

For those systems that do not show signs of azimuthal and/or orbital variation, we may simply set $A_\text{az} = A_\text{orb} = 1$ in Equations~\ref{eq:azimuthal} and \ref{eq:orbital}, thereby setting the weighting effect to 0. We show in Figure~\ref{fig:variation_gaussian} these weighting values given the particle angle or ring true anomaly for multiple values of $A$ and $\sigma$. 

When the true anomaly at the current orbital position falls between that of the turn on and off threshold values, the produced rings contribute to the visible dust plume. We should not expect, however, that the turn on or off of dust production is instantaneous, but rather that there is a gradual change in the production rate. Hence as an additional feature, we have implemented a turn on/off smoothing in the dust production that was not present in previous versions of the geometric model. To do this, we use a half-Gaussian function to decrease the weight of rings that are outside of the usual dust production regime; this function has a value of 1 at the usual dust production true anomaly bounds, and decreases with standard deviation $\sigma_\text{gradual}$ outside of the usual dust production regime. To include this we therefore have to inject some rings outside of the usual dust production true anomaly bounds, $\nuonoff$. We modify the previous bounds to now be
\begin{equation}
    \nuon = \text{max} \left\{-180^\circ, ~ \nuon - 2 \sigma_\text{gradual}\right\}; \quad \nuoff = \text{min} \left\{180^\circ, ~ \nuoff + 2 \sigma_\text{gradual}\right\}
\end{equation}
where we include rings up to two standard deviations outside of the usual bounds. For each ring, we then apply a histogram weighting of
\begin{equation}
	\delta w_\text{gradual} = \begin{cases}
	    \exp \left(\frac{(\nu_\text{ring} - \nuon)^2}{2 \sigma_\text{gradual}^2}\right) & \nu_\text{ring} < \nuon \\
        1 & \nuon \leq \nu_\text{ring} \leq \nuoff \\
        \exp \left(\frac{(\nu_\text{ring} - \nuoff)^2}{2 \sigma_\text{gradual}^2}\right) & \nuoff < \nu_\text{ring}
	\end{cases}
\end{equation}
to emulate a gradual dust production turn on and off. With this as the final required weighting factor, the overall weighting on each particle in the point cloud is calculated as
\begin{equation}
    w = \delta w_\text{az} \times \delta w_\text{orb} \times \delta w_\text{gradual} \label{eq:weighting}
\end{equation}
For the Apep system in particular, we make more modifications to the total weighting of each particle which we discuss in Section~\ref{sec:tertiary}. 

\begin{figure}
    \centering
    \includegraphics[width=0.65\linewidth]{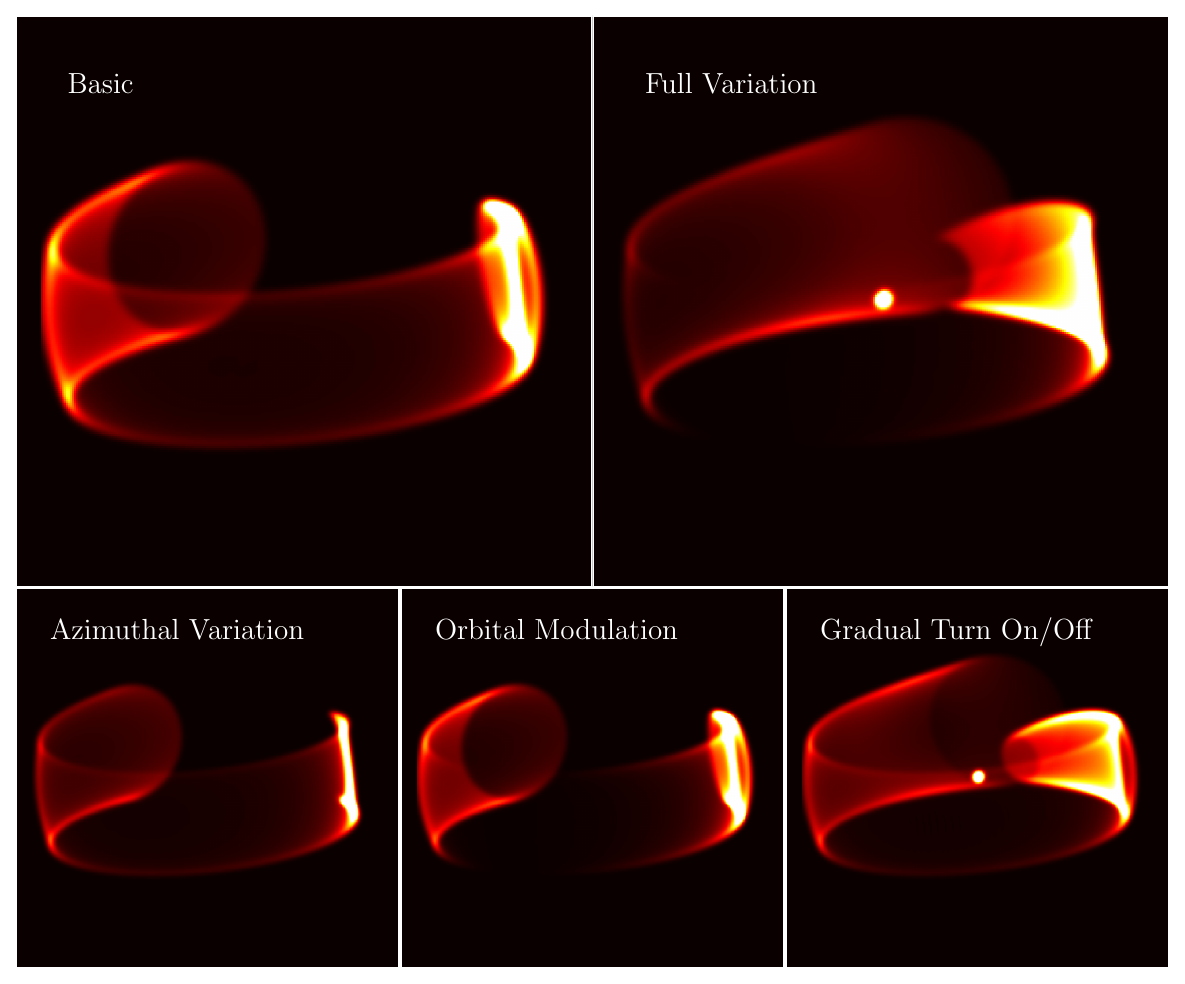}
    \caption{Each higher order variation affects the dust plume in different ways. The top right plot shows how an arbitrary test system appears with only the `basic' model implementation as described in Section~\ref{sec:ring_model}. The bottom row shows each of the higher order dust production variations discussed so far, and their effect on the `basic' plume in isolation. The top right plot shows the effect of including all of these features at once. The value of each effect in the figure is: $\sigma_\text{az} = 60^\circ$; $A_\text{az} = -1$; $\sigma_\text{orb} = 40^\circ$; $A_\text{orb} = 0$; $\sigma_\text{gradual} = 19^\circ$.}
    \label{fig:variation_effects}
\end{figure}

\subsection{Acceleration in the Dust Shell} \label{sec:dust-accel}
Wolf-Rayet stars are among the most luminous stellar objects, and so the environment immediately surrounding them is engulfed in intense radiation. The winds of these stars are known to accelerate to a terminal velocity some time after leaving the photosphere of the star because of the intense radiation pressure from the star itself \citep{Abbott1978ApJ}. Further downstream, we should expect that the dust grains formed would experience intense radiation pressure given that WR stars are at a significant fraction of their Eddington luminosity \citep{Grafener2011A&A, Maeder2012A&A}. This radiation pressure has been observed accelerating gas around clusters of stars \citep{Murray2011ApJ}, and has been used to explain the apparent acceleration of dust shells around WR\,140 \citep{Han2022Natur}. 

In the paper describing the radiative acceleration of dust around WR\,140, the authors do not explicitly include acceleration into their geometric model. For the first time, we include this acceleration in our model. \citet{Han2022Natur} propose an acceleration model that is largely physically motivated: the dust nucleates at some distance from the star in an optically thick regime while accelerating at a constant rate, then becomes optically thin some time later and proceeds to accelerate at a decreasing rate. The strength of acceleration in this model is parameterised by radial distance from the central WR+O binary, as one would expect from radiation pressure which follows an inverse-square law. Unfortunately elements of this prescription are difficult to integrate into our geometric model; the ejected rings are parameterised mainly by their age, which is subsequently used to calculate their (non-accelerated) radial distance. In order to incorporate the acceleration terms from \citet{Han2022Natur} into our geometric model, an iterative method would be required to find the correct accelerated distance. This introduces computational complexity and is at odds with gradient computation in \textsc{Jax}. 

To maximise computational efficiency and maintain gradient calculations, we instead use a phenomenological model for acceleration parameterised by the dust ring age: the velocity of each ring is the exponential decay of some initial velocity $v_i$ to a terminal velocity $v_\infty$ in the radial direction from the central WR binary. That is,
\begin{equation}
    v(t_\text{age}) = v_\infty + (v_i - v_\infty ) \, \exp \left(-10^\mathcal{A} \, t_\text{age}\right) \label{eq:acceleration}
\end{equation}
where $\mathcal{A}$ is a scaling constant which determines the slope of the exponential (i.e. the strength of the acceleration). When $v$ is in units of km\,s$^{-1}$ and $t_\text{age}$ in units of years, we find this exponential constant should be roughly in the range $10^{-5}$ to $10^0$~years$^{-1}$, or $-5 \lesssim \mathcal{A} \lesssim 0$. 

Including this acceleration into the model means that we replace the $v_\text{dust}$ parameter in Equation~\ref{eq:ring_expand} so that the Equation now reads
\begin{equation}
    \vec{r}_\text{particle} = v(t_\text{age}) \, t_\text{age} \, \mathbf{\hat{r}}_\text{particle} \label{eq:ring_expand_accel}
\end{equation}

In modelling acceleration this way, we note that we can also easily simulate dust deceleration as needed. In Equation~\ref{eq:acceleration} we see that when $v_\infty < v_i$, the rings are decelerating towards a terminal velocity; conversely, $v_\infty > v_i$ results in dust ring acceleration as we would model in WR\,140, for example. The detailed consequences of modelling acceleration with an exponential instead of a power law have not been exhaustively explored. However the acceleration term is dominant only at early orbital phases when the dust is close to the luminous stars. At present there does not exist a large body of observations of the CWB systems at these phases, making comparison difficult. Since \cite{Han2022Natur} and our implementation of acceleration both yield velocity converging to a terminal value, we expect that any difference in geometric fits would be small and not the dominant source of model error.

\subsection{Modelling Wind Anisotropy} \label{sec:anisotropy}

One of the biggest unanswered questions about Wolf-Rayet colliding wind binaries is whether their spiral nebulae can exhibit evidence of wind anisotropy, particularly in the case of the Apep system \citep{Callingham2019NatAs}. The dominant wind in the Apep system has been suggested as being anisotropic on account of the large discrepancy between the polar wind speed (measured through spectroscopy) and the nebula expansion speed (measured kinematically); the idea here is that WC star in Apep may harbour a slow, but dense equatorial wind which is responsible for the visible dust plume as well as a fast and sparse polar wind. This interpretation suggests that this star is critically rotating, in which case Apep is the most likely LGRB progenitor candidate known in our Galaxy. It is therefore essential to look for any evidence of wind anisotropy in the visible plume to investigate this hypothesis.

\begin{figure}
    \centering
    \resizebox{.85\textwidth}{!}{
        \input{anisotropy.tikz}\hspace{1cm}
        \input{anisotropy_top.tikz}
    }
    \caption{Wind anisotropies may be visible in the dust plume if the wind-dominant star is inclined with respect to the orbital plane (left). In principle, the axis of rotation could also be misaligned from the periapsis of orbit (right) which might affect the wind anisotropy evidence in elliptical orbits.
    We have an image available \externalgithublink{effect of spin-orbit misaligned wind anisotropy}{https://github.com/ryanwhite1/ryanwhite1.github.io/blob/8933b7b0cd479cf35d0d7ee51b59a54c255d0132/permanent_storage/apep_photos/Anisotropy_Effects.png} that shows the effect of a modest ($\sim 30^\circ$) spin-orbit misalignment with a faster polar wind.}
    \label{fig:anisotropy_schematic}
\end{figure}
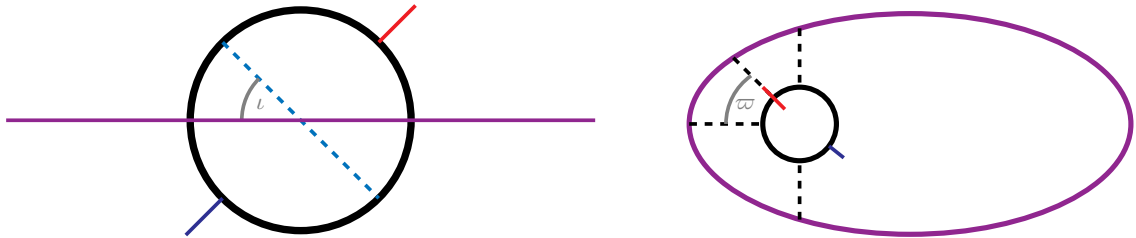

We present here the first attempt at integrating an anisotropic wind into the geometric model for the Wolf-Rayet CWBs. Evidence of anisotropy in the dust plume may be present if the star with the higher momentum wind is inclined with respect to the orbital plane of the binary; this would mean that the wind at the shock interface would change in momentum depending on the orbital position of the stars. The inclination of this star may also be misaligned with the argument of periastron of the orbit, $\omega$, which would affect the anisotropy-true anomaly correlation. These ideas are shown in Figure~\ref{fig:anisotropy_schematic}. 

We parameterise the strength of wind anisotropy based on the origin latitude of the wind that is currently causing the colliding wind shock, $\vartheta$. We take $\vartheta = 0$ as being along the equator of the windy star, and so the value of $\vartheta$ at any point along the orbit is calculated as 
\begin{equation}
    \vartheta (\nu) = |\iota \sin (\nu - \varpi) \label{eq:latitude}|
\end{equation}
where $\iota$ is the inclination of the star with respect to the orbital plane, and $\varpi$ is the argument of periastron offset, and the absolute value assumes that the wind anisotropy is symmetric about the equator. 

We expect that a wind anisotropy would manifest itself as a true anomaly dependence on the plume opening angle and/or expansion velocity in the CWB nebula. This is because if the mass-loss or speed is latitude-dependent, the momentum ratio of the winds at the shock interface would change depending on the wind origin and hence the plume geometry should change as a result. We considered implementing a model in which the rings were elliptical (as opposed to circular), but the WCR is always such a distance from the two stars that the local wind field is essentially planar. 

Without detailed knowledge of the exact 3D wind fields around these WC/WN stars, we phenomenologically model these opening angle and expansion velocity perturbations due to anisotropy as multipliers onto the standard parameters, much in the same way we modelled azimuthal and orbital variations as histogram weighting multipliers. These multipliers are 
\begin{align}
    \delta v(\vartheta) &= 1 + \left(\frac{v_\text{polar}}{v(t_\text{age})} - 1\right) \tanh \left(10^{m_v} \vartheta^{p_v}\right) \\
    \delta \theta_\text{OA}(\vartheta) &= 1 + \left(\frac{\theta_\text{OA, polar}}{\theta_\text{OA}} - 1\right) \tanh \left(10^{m_\text{OA}} \vartheta^{p_\text{OA}}\right) 
\end{align}
where we have three additional parameters for each of the velocity and opening angle multipliers: the parameter of subscript `polar' denotes the value of that parameter due to the polar wind, the $m_v$ and $m_\text{OA}$ represent the value of a constant multiplier onto the latitude dependence of the wind, and $p_v$/$p_\text{OA}$ impose a power law dependence of the anisotropy strength onto the latitude. The $\tanh$ function was used here as this allows us to sensitively change the slope and shape of the multiplier dependence on $\phi$. As a caveat, this manifests only as a monotonically changing multiplier value and cannot represent a latitude dependence with turning points. This non-monotonic behaviour has been seen in some numerical models of massive stellar winds \citep{Hastings2023A&A}, although it is common to assume smoothly changing monotonic wind parameters vs latitude for evolved massive stars \citep{van-Marle2008A&A}. The true WR wind morphology, including the effects of rotation, would require detailed stellar evolution and magnetohydrodynamical models, and this does not appear to have been published in the literature \citep[although this has been recently done for non-rotating WR stars in][]{Moens2022A&A}.

\subsection{Miscellaneous Features and Details} \label{sec:nonphysicalfeatures}
There exist a number of features we have implemented into the geometric model that have either little or no effect on the physical geometry. In the aim of performing statistical inference with our model, we have added in two parameters that affect only the pixel values of the final render; the first is a brightness ceiling on the rendered array as an emulation for image pixel saturation, and the second is a power law scaling of the remaining pixels mainly to exaggerate otherwise dim pixels. Allowing for these parameters to change makes it easier to compare simulated images to real data, and may allow for faster convergence in statistical methods on our simulations. 

The next modification made to the geometric model pertains to the azimuthal angles of particles within each generated ring. Equation~\ref{eq:ring_particles} describes the initial Cartesian coordinates of particles within each ring, parameterised by some azimuthal angle $\theta_\text{particle}$ unique to each particle within the ring. In early versions of our model, the grid of these azimuthal angles for particles within each ring was identical across all of the rings in the plume; what we saw was that this accentuated the quantisation error of approximating these continuous rings as a series of particles. Since the particles were at identical angular positions within neighbouring rings, `stripes' of nebulosity would be visible in the final image. One solution to this would be to simply increase the number of particles in each ring, although this would correspondingly increase compute time. To minimise this effect, we instead shift the angular coordinate of all particles within a ring by $1$ radian across each consecutive ring created. This imposes some quasi-randomness in the position of the particles across neighbouring rings, especially since the angular coordinates are mod $2\pi$. 

Since we calculate the position of each star at all times in the model, it is relatively straightforward to take this position and plot the stars on the final rendered image. To do this, we generate the simulated image of the colliding wind nebula, then inject Gaussian flux profiles on the angular positions of the stars at that epoch. The user can determine the amplitude of the flux (how bright the star is), as well as the standard deviation of the flux profile (how much the flux bleeds into neighbouring pixels). Including the stars in each image helps to anchor the geometry of the plume with respect to its origin, and the position of stars becomes particularly important in our discussion in Section~\ref{sec:tertiary}. 

The final small addition to the geometric model that we describe is the dust nucleation distance. Dust is expected to nucleate at some distance, $r_\text{nuc}$, behind the colliding wind shock \citep[referred to as `dust-forming separation' in][for example]{Eatson2022MNRASa} where the conditions are favourable. This nucleation distance is estimated to be typically on the order of tens of au for the WR CWBs, sufficiently far downstream for both mixing of the two winds and attenuation of the stars' ionising radiation. Because this nucleation distance is relatively small, the gaps in the models due to dust not yet nucleated are limited to phases close to periastron for the episodic dust producers. To account for this nucleation distance in the model, we include another ring weighting multiplier in the form of a heaviside step function
\begin{equation}
    \delta w_\text{nuc} = \begin{cases}
        1 &  ||\vec{r}_\text{particle}|| \geq r_\text{nuc} \\
        0 &  \text{otherwise}
    \end{cases}
\end{equation}
which is then included onto Equation~\ref{eq:weighting}. Including this feature mainly results in a delay in dust production from the orbital motion of the stars, and has no impact on dust shells that are not actively forming dust. The strongest evidence for delayed onset of dust formation, apart from in hydrodynamical simulations, comes from CWB light curves where phase shifts between orbital data and infrared peaks are apparent.

\subsection{Modelling of Effects from a Tertiary Companion} \label{sec:tertiary}

Previous models of Apep have been successful in reproducing most but not all of the geometry of the colliding wind nebula. In particular, the northern region of the VISIR image pane of Figure~\ref{fig:apep-cone} is not suitably reproduced by previous versions of the geometric code (see the top pane of Figure~\ref{fig:apep-cone} for a representative example of previous attempts). Upon realising that the cavity in the plume aligned perfectly with the apparent position of the northern companion of Apep's inner binary, we began investigating how a distant third massive companion could influence a colliding wind nebula. The dominant effect appears to be destruction of the already formed dust once it collides with the tertiary O star wind. With this in mind, our model `destroys' dust that falls within some angular distance of the tertiary star with respect to the inner WR+WR binary. 

Like the generation of the colliding wind nebula, the cavity region predominantly relies on a few key parameters: the position of the tertiary star with respect to the inner binary and the cone opening angle of the cavity. For the former, in the code we define the position of the third star in spherical coordinates with respect to the inner binary orbital plane. That is, we choose the angular location of the cavity with a polar and azimuthal angle, $\beta_\text{tert}$ and $\alpha_\text{tert}$, as well as the radial distance of the third star, $r_\text{tert}$.

With these parameters, we de-weight the generated points whose angular coordinates are close enough to the centre of the cone projection axis. To calculate the angular distance, we use the formula from \citep{Kells1940}, modified to fit with the convention that the inclination $\theta = 0$ is at the north pole rather than $\theta = -\pi/2$:
\begin{align}
    \Theta = \arccos &\left[\cos (\beta_\text{tert}) \cos (\beta_\text{part}) + \sin (\beta_\text{tert}) \sin (\beta_\text{part}) \cos (\alpha_\text{tert} - \alpha_\text{part}) \right]
\end{align}
We then change the weighting of each point in the rendered image by a Gaussian according to their angular distance from the projected cone,
\begin{equation}
    \delta w_\text{tert} = 1 - A_\text{tert} \exp \left(-\frac{2\Theta^2}{\theta_\text{OA,tert}^2}\right) \label{eq:tertiary_gauss}
\end{equation}
where $A_\text{tert}$ is the amplitude of the dust destruction, $\Theta$ is the angular distance for each particle from the centre of the cone, and $\theta_\text{OA,tert}$ is the cavity opening angle. This is then appended onto Equation~\ref{eq:weighting} as another pre-processing step onto the final rendered plume. Rather than setting $\delta w_\text{tert} = 0$ for all particles with $\Theta < \theta_\text{OA,tert}$ we chose to model this effect with a Gaussian fall off; we would expect that all dust particles having a trajectory that impacts the tertiary O star should be destroyed, but as their impact parameter gets smaller (that is, as their trajectory deviates further from the tertiary star's position) we expect that proportionally fewer dust grains will be destroyed. As shown in Figure~\ref{fig:apep-fit} (which includes the destruction effect with parameters as in Table~\ref{tab:apep-photodissociation}), this reproduces the geometry around the tertiary star well.  

While this phenomenological prescription was motivated by the Apep system, it can, in principle, be applied to other hierarchical triples hosting WR-CWBs. Unfortunately, WR-CWBs that can be resolved with directly imaged are exceedingly rare with only a handful in the Galaxy. Hierarchical triple variants of these, especially those whose tertiary companion is close enough to sculpt surviving and bright dust, are rarer still. At present, there appear to be only two other confirmed hierarchical triple CWBs hosting WR stars: WR\,104 \citep{Wallace2002ASPC, Soulain2018A&A} and WR\,147 \citep{Rodriguez2020ApJ}. For the former, the tertiary star seems too distant to affect the spiral dust plume, and radio imagery of the latter implies that the tertiary shock does not affect the spiral nebula. Still, this cavity prescription may be useful in later studies should similar nebular morphology be observed in other systems.

\bibliography{references}{}
\bibliographystyle{aasjournalv7}

\end{document}

%% file: compass_and_scale.tikz
\begin{tikzpicture}

    


    \draw[line width=0.5mm, ->, white] (8.0126, 2) -- (7.95, 2.998) node[above]{N};
    \draw[line width=0.5mm, ->, white] (8.2, 2.2126) -- (7.202, 2.150) node[left]{E};

    \draw[line width=0.5mm, white] (-6, -8) -- (-4.7, -8) node[midway, above]{24000\,au};

\end{tikzpicture}

%% file: anisotropy.tikz
\begin{tikzpicture}
    \draw[line width=1mm] (0, 0) circle (1.5cm);

    \draw[line width=0.5mm, dashed, RoyalBlue] (-1.06, 1.06) -- (1.06, -1.06);
    \draw[line width=0.5mm, Red] (1.06, 1.06) -- (1.56, 1.56);
    \draw[line width=0.5mm, Blue] (-1.06, -1.06) -- (-1.56, -1.56);
    \draw[line width=0.5mm, gray] (-0.8, 0) arc (180:135:0.8) node [midway, right, yshift=-0.8mm] {$\iota$};
    
    \draw[line width=0.5mm, Plum] (-4, 0) -- (4, 0);

\end{tikzpicture}

%% file: anisotropy_top.tikz
\begin{tikzpicture}
    \draw[line width=0.7mm] (-1.5, 0) circle (0.5cm);
    \draw[line width=0.7mm, Plum] (0,0) ellipse (3cm and 1.5cm);

    \draw[line width=0.5mm, dashed] (-1.5, 1.3) -- (-1.5, 0.5);
    \draw[line width=0.5mm, dashed] (-1.5, -1.3) -- (-1.5, -0.5);

    \draw[line width=0.5mm, dashed] (-3, 0) -- (-2, 0);
    \draw[line width=0.5mm, dashed] (-2.4, 0.9) -- (-1.9, 0.4);
    \draw[line width=0.5mm, gray] (-2.5, 0) arc (180:125:0.8) node [midway, right, yshift=-1mm, xshift=-1mm] {$\varpi$};
    
    \draw[line width=0.5mm, Red] (-1.7, 0.2) -- (-2, 0.5);
    \draw[line width=0.5mm, Blue] (-1.1, -0.3) -- (-0.9, -0.46);

    

\end{tikzpicture}

%% file: main.bbl
\begin{thebibliography}{}
\expandafter\ifx\csname natexlab\endcsname\relax\def\natexlab#1{#1}\fi
\providecommand{\url}[1]{\href{#1}{#1}}
\providecommand{\dodoi}[1]{doi:~\href{http://doi.org/#1}{\nolinkurl{#1}}}
\providecommand{\doeprint}[1]{\href{http://ascl.net/#1}{\nolinkurl{http://ascl.net/#1}}}
\providecommand{\doarXiv}[1]{\href{https://arxiv.org/abs/#1}{\nolinkurl{https://arxiv.org/abs/#1}}}

\bibitem[{D.~C. {Abbott}(1978){Abbott}}]{Abbott1978ApJ}
{Abbott}, D.~C. 1978, \bibinfo{title}{{The terminal velocities of stellar winds from early-type stars.},} \apj, 225, 893, \dodoi{10.1086/156554}

\bibitem[{J.~M.~O. {Antognini}(2015){Antognini}}]{Antognini2015MNRAS}
{Antognini}, J.~M.~O. 2015, \bibinfo{title}{{Timescales of Kozai-Lidov oscillations at quadrupole and octupole order in the test particle limit},} \mnras, 452, 3610, \dodoi{10.1093/mnras/stv1552}

\bibitem[{J.~M.~O. {Antognini} \& T.~A. {Thompson}(2016){Antognini} \& {Thompson}}]{Antognini2016MNRAS}
{Antognini}, J. M.~O., \& {Thompson}, T.~A. 2016, \bibinfo{title}{{Dynamical formation and scattering of hierarchical triples: cross-sections, Kozai-Lidov oscillations, and collisions},} \mnras, 456, 4219, \dodoi{10.1093/mnras/stv2938}

\bibitem[{ {Astropy Collaboration} {et~al.}(2013){Astropy Collaboration}, {Robitaille}, {Tollerud}, {Greenfield}, {Droettboom}, {Bray}, {Aldcroft}, {Davis}, {Ginsburg}, {Price-Whelan}, {Kerzendorf}, {Conley}, {Crighton}, {Barbary}, {Muna}, {Ferguson}, {Grollier}, {Parikh}, {Nair}, {Unther}, {Deil}, {Woillez}, {Conseil}, {Kramer}, {Turner}, {Singer}, {Fox}, {Weaver}, {Zabalza}, {Edwards}, {Azalee Bostroem}, {Burke}, {Casey}, {Crawford}, {Dencheva}, {Ely}, {Jenness}, {Labrie}, {Lim}, {Pierfederici}, {Pontzen}, {Ptak}, {Refsdal}, {Servillat}, \& {Streicher}}]{Astropy2013A&A}
{Astropy Collaboration}, {Robitaille}, T.~P., {Tollerud}, E.~J., {et~al.} 2013, \bibinfo{title}{{Astropy: A community Python package for astronomy},} \aap, 558, A33, \dodoi{10.1051/0004-6361/201322068}

\bibitem[{ {Astropy Collaboration} {et~al.}(2018){Astropy Collaboration}, {Price-Whelan}, {Sip{\H{o}}cz}, {G{\"u}nther}, {Lim}, {Crawford}, {Conseil}, {Shupe}, {Craig}, {Dencheva}, {Ginsburg}, {VanderPlas}, {Bradley}, {P{\'e}rez-Su{\'a}rez}, {de Val-Borro}, {Aldcroft}, {Cruz}, {Robitaille}, {Tollerud}, {Ardelean}, {Babej}, {Bach}, {Bachetti}, {Bakanov}, {Bamford}, {Barentsen}, {Barmby}, {Baumbach}, {Berry}, {Biscani}, {Boquien}, {Bostroem}, {Bouma}, {Brammer}, {Bray}, {Breytenbach}, {Buddelmeijer}, {Burke}, {Calderone}, {Cano Rodr{\'\i}guez}, {Cara}, {Cardoso}, {Cheedella}, {Copin}, {Corrales}, {Crichton}, {D'Avella}, {Deil}, {Depagne}, {Dietrich}, {Donath}, {Droettboom}, {Earl}, {Erben}, {Fabbro}, {Ferreira}, {Finethy}, {Fox}, {Garrison}, {Gibbons}, {Goldstein}, {Gommers}, {Greco}, {Greenfield}, {Groener}, {Grollier}, {Hagen}, {Hirst}, {Homeier}, {Horton}, {Hosseinzadeh}, {Hu}, {Hunkeler}, {Ivezi{\'c}}, {Jain}, {Jenness}, {Kanarek}, {Kendrew}, {Kern}, {Kerzendorf}, {Khvalko}, {King}, {Kirkby}, {Kulkarni},
  {Kumar}, {Lee}, {Lenz}, {Littlefair}, {Ma}, {Macleod}, {Mastropietro}, {McCully}, {Montagnac}, {Morris}, {Mueller}, {Mumford}, {Muna}, {Murphy}, {Nelson}, {Nguyen}, {Ninan}, {N{\"o}the}, {Ogaz}, {Oh}, {Parejko}, {Parley}, {Pascual}, {Patil}, {Patil}, {Plunkett}, {Prochaska}, {Rastogi}, {Reddy Janga}, {Sabater}, {Sakurikar}, {Seifert}, {Sherbert}, {Sherwood-Taylor}, {Shih}, {Sick}, {Silbiger}, {Singanamalla}, {Singer}, {Sladen}, {Sooley}, {Sornarajah}, {Streicher}, {Teuben}, {Thomas}, {Tremblay}, {Turner}, {Terr{\'o}n}, {van Kerkwijk}, {de la Vega}, {Watkins}, {Weaver}, {Whitmore}, {Woillez}, {Zabalza}, \& {Astropy Contributors}}]{Astropy2018AJ}
{Astropy Collaboration}, {Price-Whelan}, A.~M., {Sip{\H{o}}cz}, B.~M., {et~al.} 2018, \bibinfo{title}{{The Astropy Project: Building an Open-science Project and Status of the v2.0 Core Package},} \aj, 156, 123, \dodoi{10.3847/1538-3881/aabc4f}

\bibitem[{ {Astropy Collaboration} {et~al.}(2022){Astropy Collaboration}, {Price-Whelan}, {Lim}, {Earl}, {Starkman}, {Bradley}, {Shupe}, {Patil}, {Corrales}, {Brasseur}, {N{\"o}the}, {Donath}, {Tollerud}, {Morris}, {Ginsburg}, {Vaher}, {Weaver}, {Tocknell}, {Jamieson}, {van Kerkwijk}, {Robitaille}, {Merry}, {Bachetti}, {G{\"u}nther}, {Aldcroft}, {Alvarado-Montes}, {Archibald}, {B{\'o}di}, {Bapat}, {Barentsen}, {Baz{\'a}n}, {Biswas}, {Boquien}, {Burke}, {Cara}, {Cara}, {Conroy}, {Conseil}, {Craig}, {Cross}, {Cruz}, {D'Eugenio}, {Dencheva}, {Devillepoix}, {Dietrich}, {Eigenbrot}, {Erben}, {Ferreira}, {Foreman-Mackey}, {Fox}, {Freij}, {Garg}, {Geda}, {Glattly}, {Gondhalekar}, {Gordon}, {Grant}, {Greenfield}, {Groener}, {Guest}, {Gurovich}, {Handberg}, {Hart}, {Hatfield-Dodds}, {Homeier}, {Hosseinzadeh}, {Jenness}, {Jones}, {Joseph}, {Kalmbach}, {Karamehmetoglu}, {Ka{\l}uszy{\'n}ski}, {Kelley}, {Kern}, {Kerzendorf}, {Koch}, {Kulumani}, {Lee}, {Ly}, {Ma}, {MacBride}, {Maljaars}, {Muna}, {Murphy}, {Norman},
  {O'Steen}, {Oman}, {Pacifici}, {Pascual}, {Pascual-Granado}, {Patil}, {Perren}, {Pickering}, {Rastogi}, {Roulston}, {Ryan}, {Rykoff}, {Sabater}, {Sakurikar}, {Salgado}, {Sanghi}, {Saunders}, {Savchenko}, {Schwardt}, {Seifert-Eckert}, {Shih}, {Jain}, {Shukla}, {Sick}, {Simpson}, {Singanamalla}, {Singer}, {Singhal}, {Sinha}, {Sip{\H{o}}cz}, {Spitler}, {Stansby}, {Streicher}, {{\v{S}}umak}, {Swinbank}, {Taranu}, {Tewary}, {Tremblay}, {de Val-Borro}, {Van Kooten}, {Vasovi{\'c}}, {Verma}, {de Miranda Cardoso}, {Williams}, {Wilson}, {Winkel}, {Wood-Vasey}, {Xue}, {Yoachim}, {Zhang}, {Zonca}, \& {Astropy Project Contributors}}]{Astropy2022ApJ}
{Astropy Collaboration}, {Price-Whelan}, A.~M., {Lim}, P.~L., {et~al.} 2022, \bibinfo{title}{{The Astropy Project: Sustaining and Growing a Community-oriented Open-source Project and the Latest Major Release (v5.0) of the Core Package},} \apj, 935, 167, \dodoi{10.3847/1538-4357/ac7c74}

\bibitem[{S. {Bloot} {et~al.}(2022){Bloot}, {Callingham}, \& {Marcote}}]{Bloot2022MNRAS}
{Bloot}, S., {Callingham}, J.~R., \& {Marcote}, B. 2022, \bibinfo{title}{{Radio modelling of the brightest and most luminous non-thermal colliding-wind binary Apep},} \mnras, 509, 475, \dodoi{10.1093/mnras/stab2976}

\bibitem[{J. Bradbury {et~al.}(2018)Bradbury, Frostig, Hawkins, Johnson, Leary, Maclaurin, Necula, Paszke, Vander{P}las, Wanderman-{M}ilne, \& Zhang}]{jax2018github}
Bradbury, J., Frostig, R., Hawkins, P., {et~al.} 2018, {JAX}: composable transformations of {P}ython+{N}um{P}y programs, 0.3.13 \url{http://github.com/google/jax}

\bibitem[{J.~R. {Callingham} {et~al.}(2020){Callingham}, {Crowther}, {Williams}, {Tuthill}, {Han}, {Pope}, \& {Marcote}}]{Callingham2020MNRAS}
{Callingham}, J.~R., {Crowther}, P.~A., {Williams}, P.~M., {et~al.} 2020, \bibinfo{title}{{Two Wolf-Rayet stars at the heart of colliding-wind binary Apep},} \mnras, 495, 3323, \dodoi{10.1093/mnras/staa1244}

\bibitem[{J.~R. {Callingham} {et~al.}(2019){Callingham}, {Tuthill}, {Pope}, {Williams}, {Crowther}, {Edwards}, {Norris}, \& {Kedziora-Chudczer}}]{Callingham2019NatAs}
{Callingham}, J.~R., {Tuthill}, P.~G., {Pope}, B.~J.~S., {et~al.} 2019, \bibinfo{title}{{Anisotropic winds in a Wolf-Rayet binary identify a potential gamma-ray burst progenitor},} Nature Astronomy, 3, 82, \dodoi{10.1038/s41550-018-0617-7}

\bibitem[{J. {Cant{\'o}} {et~al.}(1996){Cant{\'o}}, {Raga}, \& {Wilkin}}]{Canto1996ApJ}
{Cant{\'o}}, J., {Raga}, A.~C., \& {Wilkin}, F.~P. 1996, \bibinfo{title}{{Exact, Algebraic Solutions of the Thin-Shell Two-Wind Interaction Problem},} \apj, 469, 729, \dodoi{10.1086/177820}

\bibitem[{P.~A. {Crowther} \& C.~J. {Evans}(2009){Crowther} \& {Evans}}]{Crowther2009A&A}
{Crowther}, P.~A., \& {Evans}, C.~J. 2009, \bibinfo{title}{{A FEROS spectroscopic study of the extreme O supergiant He 3-759},} \aap, 503, 985, \dodoi{10.1051/0004-6361/200912631}

\bibitem[{S. {\lowercase{D}el Palacio} {et~al.}(2023){\lowercase{D}el Palacio}, {Garc{\'\i}a}, {De Becker}, {Altamirano}, {Bosch-Ramon}, {Benaglia}, {Marcote}, \& {Romero}}]{del-Palacio2023A&A}
{\lowercase{D}el Palacio}, S., {Garc{\'\i}a}, F., {De Becker}, M., {et~al.} 2023, \bibinfo{title}{{Evidence for non-thermal X-ray emission from the double Wolf-Rayet colliding-wind binary Apep},} \aap, 672, A109, \dodoi{10.1051/0004-6361/202245505}

\bibitem[{L.~N. {Driessen} {et~al.}(2024){Driessen}, {Pritchard}, {Murphy}, {Heald}, {Robrade}, {Das}, {Duchesne}, {Kaplan}, {Lenc}, {Lynch}, {Mitchell-Bolton}, {Pope}, {Rose}, {Stelzer}, {Wang}, \& {Zic}}]{Driessen2024PASA}
{Driessen}, L.~N., {Pritchard}, J., {Murphy}, T., {et~al.} 2024, \bibinfo{title}{{The Sydney Radio Star Catalogue: Properties of radio stars at megahertz to gigahertz frequencies},} \pasa, 41, e084, \dodoi{10.1017/pasa.2024.72}

\bibitem[{J.~W. {Eatson} {et~al.}(2022{\natexlab{a}}){Eatson}, {Pittard}, \& {Van Loo}}]{Eatson2022MNRASb}
{Eatson}, J.~W., {Pittard}, J.~M., \& {Van Loo}, S. 2022{\natexlab{a}}, \bibinfo{title}{{Exploring dust growth in the episodic WCd system WR140},} \mnras, 517, 4705, \dodoi{10.1093/mnras/stac3000}

\bibitem[{J.~W. {Eatson} {et~al.}(2022{\natexlab{b}}){Eatson}, {Pittard}, \& {Van Loo}}]{Eatson2022MNRASa}
{Eatson}, J.~W., {Pittard}, J.~M., \& {Van Loo}, S. 2022{\natexlab{b}}, \bibinfo{title}{{An exploration of dust grain growth within WCd systems using an advected scalar dust model},} \mnras, 516, 6132, \dodoi{10.1093/mnras/stac2617}

\bibitem[{K.~G. {Gayley}(2009){Gayley}}]{Gayley2009ApJ}
{Gayley}, K.~G. 2009, \bibinfo{title}{{Asymptotic Opening Angles for Colliding-Wind Bow Shocks: The Characteristic-Angle Approximation},} \apj, 703, 89, \dodoi{10.1088/0004-637X/703/1/89}

\bibitem[{G. {Gr{\"a}fener} {et~al.}(2011){Gr{\"a}fener}, {Vink}, {de Koter}, \& {Langer}}]{Grafener2011A&A}
{Gr{\"a}fener}, G., {Vink}, J.~S., {de Koter}, A., \& {Langer}, N. 2011, \bibinfo{title}{{The Eddington factor as the key to understand the winds of the most massive stars. Evidence for a {\ensuremath{\Gamma}}-dependence of Wolf-Rayet type mass loss},} \aap, 535, A56, \dodoi{10.1051/0004-6361/201116701}

\bibitem[{Y. {Han} {et~al.}(2022){Han}, {Tuthill}, {Lau}, \& {Soulain}}]{Han2022Natur}
{Han}, Y., {Tuthill}, P.~G., {Lau}, R.~M., \& {Soulain}, A. 2022, \bibinfo{title}{{Radiation-driven acceleration in the expanding WR140 dust shell},} \nat, 610, 269, \dodoi{10.1038/s41586-022-05155-5}

\bibitem[{Y. {Han} {et~al.}(submitted){Han}, {White}, {Callingham}, {Lau}, {Pope}, {Richardson}, \& {Tuthill}}]{Han2025inprep}
{Han}, Y., {White}, R.~M.~T., {Callingham}, J.~R., {et~al.} submitted, \bibinfo{title}{{The formation and evolution of dust in the colliding-wind binary Apep revealed by JWST},} \apj

\bibitem[{Y. {Han} {et~al.}(2020){Han}, {Tuthill}, {Lau}, {Soulain}, {Callingham}, {Williams}, {Crowther}, {Pope}, \& {Marcote}}]{Han2020MNRAS}
{Han}, Y., {Tuthill}, P.~G., {Lau}, R.~M., {et~al.} 2020, \bibinfo{title}{{The extreme colliding-wind system Apep: resolved imagery of the central binary and dust plume in the infrared},} \mnras, 498, 5604, \dodoi{10.1093/mnras/staa2349}

\bibitem[{C.~R. {Harris} {et~al.}(2020){Harris}, {Millman}, {van der Walt}, {Gommers}, {Virtanen}, {Cournapeau}, {Wieser}, {Taylor}, {Berg}, {Smith}, {Kern}, {Picus}, {Hoyer}, {van Kerkwijk}, {Brett}, {Haldane}, {del R{\'\i}o}, {Wiebe}, {Peterson}, {G{\'e}rard-Marchant}, {Sheppard}, {Reddy}, {Weckesser}, {Abbasi}, {Gohlke}, \& {Oliphant}}]{Harris2020Natur}
{Harris}, C.~R., {Millman}, K.~J., {van der Walt}, S.~J., {et~al.} 2020, \bibinfo{title}{{Array programming with NumPy},} \nat, 585, 357, \dodoi{10.1038/s41586-020-2649-2}

\bibitem[{B. {Hastings} {et~al.}(2023){Hastings}, {Langer}, \& {Puls}}]{Hastings2023A&A}
{Hastings}, B., {Langer}, N., \& {Puls}, J. 2023, \bibinfo{title}{{A model of anisotropic winds from rotating stars for evolutionary calculations},} \aap, 672, A60, \dodoi{10.1051/0004-6361/202245281}

\bibitem[{S. {Hattori} {et~al.}(2024){Hattori}, {Garcia}, {Murray}, {Dong}, {Dholakia}, {Degen}, \& {Foreman-Mackey}}]{jaxoplanet}
{Hattori}, S., {Garcia}, L., {Murray}, C., {et~al.} 2024, {exoplanet-dev/jaxoplanet: Astronomical time series analysis with JAX}, v0.0.2 Zenodo, \dodoi{10.5281/zenodo.10736936}

\bibitem[{T. {Hoang} \& L.~N. {Tram}(2019){Hoang} \& {Tram}}]{Hoang2019ApJ}
{Hoang}, T., \& {Tram}, L.~N. 2019, \bibinfo{title}{{Dust Rotational Dynamics in C-shocks: Rotational Disruption of Nanoparticles by Stochastic Mechanical Torques and Spinning Dust Emission},} \apj, 877, 36, \dodoi{10.3847/1538-4357/ab1845}

\bibitem[{T. {Hoang} {et~al.}(2019){Hoang}, {Tram}, {Lee}, \& {Ahn}}]{Hoang2019NatAs}
{Hoang}, T., {Tram}, L.~N., {Lee}, H., \& {Ahn}, S.-H. 2019, \bibinfo{title}{{Rotational disruption of dust grains by radiative torques in strong radiation fields},} Nature Astronomy, 3, 766, \dodoi{10.1038/s41550-019-0763-6}

\bibitem[{D. {Huijser} {et~al.}(2015){Huijser}, {Goodman}, \& {Brewer}}]{Huijser2015arXiv}
{Huijser}, D., {Goodman}, J., \& {Brewer}, B.~J. 2015, \bibinfo{title}{{Properties of the Affine Invariant Ensemble Sampler in high dimensions},} arXiv e-prints, arXiv:1509.02230, \dodoi{10.48550/arXiv.1509.02230}

\bibitem[{J.~D. Hunter(2007)Hunter}]{matplotlib}
Hunter, J.~D. 2007, \bibinfo{title}{Matplotlib: A 2D graphics environment,} Computing in Science \& Engineering, 9, 90, \dodoi{10.1109/MCSE.2007.55}

\bibitem[{L.~M. Kells \& W.~F. Kern(1940)Kells \& Kern}]{Kells1940}
Kells, L.~M., \& Kern, W.~F. 1940, Plane and Spherical Trigonometry, 2nd edn. (New York: McGraw Hill Book Company)

\bibitem[{A. {Lamberts} {et~al.}(2012){Lamberts}, {Dubus}, {Lesur}, \& {Fromang}}]{Lamberts2012A&A}
{Lamberts}, A., {Dubus}, G., {Lesur}, G., \& {Fromang}, S. 2012, \bibinfo{title}{{Impact of orbital motion on the structure and stability of adiabatic shocks in colliding wind binaries},} \aap, 546, A60, \dodoi{10.1051/0004-6361/201219006}

\bibitem[{R.~M. {Lau} {et~al.}(2020){Lau}, {Eldridge}, {Hankins}, {Lamberts}, {Sakon}, \& {Williams}}]{Lau2020ApJa}
{Lau}, R.~M., {Eldridge}, J.~J., {Hankins}, M.~J., {et~al.} 2020, \bibinfo{title}{{Revisiting the Impact of Dust Production from Carbon-rich Wolf-Rayet Binaries},} \apj, 898, 74, \dodoi{10.3847/1538-4357/ab9cb5}

\bibitem[{R.~M. {Lau} {et~al.}(2022){Lau}, {Hankins}, {Han}, {Argyriou}, {Corcoran}, {Eldridge}, {Endo}, {Fox}, {Garcia Marin}, {Gull}, {Jones}, {Hamaguchi}, {Lamberts}, {Law}, {Madura}, {Marchenko}, {Matsuhara}, {Moffat}, {Morris}, {Morris}, {Onaka}, {Ressler}, {Richardson}, {Russell}, {Sanchez-Bermudez}, {Smith}, {Soulain}, {Stevens}, {Tuthill}, {Weigelt}, {Williams}, \& {Yamaguchi}}]{Lau2022NatAs}
{Lau}, R.~M., {Hankins}, M.~J., {Han}, Y., {et~al.} 2022, \bibinfo{title}{{Nested dust shells around the Wolf-Rayet binary WR 140 observed with JWST},} Nature Astronomy, 6, 1308, \dodoi{10.1038/s41550-022-01812-x}

\bibitem[{R.~M. {Lau} {et~al.}(2023){Lau}, {Wang}, {Hankins}, {Currie}, {Deo}, {Endo}, {Guyon}, {Han}, {Jones}, {Jovanovic}, {Lozi}, {Moffat}, {Onaka}, {Ruane}, {Sander}, {Tinyanont}, {Tuthill}, {Weigelt}, {Williams}, \& {Vievard}}]{Lau2023ApJ}
{Lau}, R.~M., {Wang}, J., {Hankins}, M.~J., {et~al.} 2023, \bibinfo{title}{{From Dust to Nanodust: Resolving Circumstellar Dust from the Colliding-wind Binary Wolf-Rayet 140},} \apj, 951, 89, \dodoi{10.3847/1538-4357/acd4c5}

\bibitem[{M.~N. {Lemaster} {et~al.}(2007){Lemaster}, {Stone}, \& {Gardiner}}]{Lemaster2007ApJ}
{Lemaster}, M.~N., {Stone}, J.~M., \& {Gardiner}, T.~A. 2007, \bibinfo{title}{{Effect of the Coriolis Force on the Hydrodynamics of Colliding-Wind Binaries},} \apj, 662, 582, \dodoi{10.1086/515431}

\bibitem[{E.~P. {Lieb} {et~al.}(2025){Lieb}, {Lau}, {Hoffman}, {Corcoran}, {Garcia Marin}, {Gull}, {Hamaguchi}, {Han}, {Hankins}, {Jones}, {Madura}, {Marchenko}, {Matsuhara}, {Millour}, {Moffat}, {Morris}, {Morris}, {Onaka}, {Perrin}, {Rest}, {Richardson}, {Russell}, {Sanchez-Bermudez}, {Soulain}, {Tuthill}, {Weigelt}, \& {Williams}}]{Lieb2025ApJ}
{Lieb}, E.~P., {Lau}, R.~M., {Hoffman}, J.~L., {et~al.} 2025, \bibinfo{title}{{Dynamic Imprints of Colliding-wind Dust Formation from WR 140},} \apjl, 979, L3, \dodoi{10.3847/2041-8213/ad9aa9}

\bibitem[{A. {Maeder} {et~al.}(2012){Maeder}, {Georgy}, {Meynet}, \& {Ekstr{\"o}m}}]{Maeder2012A&A}
{Maeder}, A., {Georgy}, C., {Meynet}, G., \& {Ekstr{\"o}m}, S. 2012, \bibinfo{title}{{On the Eddington limit and Wolf-Rayet stars},} \aap, 539, A110, \dodoi{10.1051/0004-6361/201118328}

\bibitem[{B. {Marcote} {et~al.}(2021){Marcote}, {Callingham}, {De Becker}, {Edwards}, {Han}, {Schulz}, {Stevens}, \& {Tuthill}}]{Marcote2021MNRAS}
{Marcote}, B., {Callingham}, J.~R., {De Becker}, M., {et~al.} 2021, \bibinfo{title}{{AU-scale radio imaging of the wind collision region in the brightest and most luminous non-thermal colliding wind binary Apep},} \mnras, 501, 2478, \dodoi{10.1093/mnras/staa3863}

\bibitem[{F.~L. {Markley}(1995){Markley}}]{Markley1995CeMDA}
{Markley}, F.~L. 1995, \bibinfo{title}{{Kepler Equation Solver},} Celestial Mechanics and Dynamical Astronomy, 63, 101, \dodoi{10.1007/BF00691917}

\bibitem[{N. {Moens} {et~al.}(2022){Moens}, {Poniatowski}, {Hennicker}, {Sundqvist}, {El Mellah}, \& {Kee}}]{Moens2022A&A}
{Moens}, N., {Poniatowski}, L.~G., {Hennicker}, L., {et~al.} 2022, \bibinfo{title}{{First 3D radiation-hydrodynamic simulations of Wolf-Rayet winds},} \aap, 665, A42, \dodoi{10.1051/0004-6361/202243451}

\bibitem[{T. {Moore} {et~al.}(2023){Moore}, {Smartt}, {Nicholl}, {Srivastav}, {Stevance}, {Jess}, {Grant}, {Fulton}, {Rhodes}, {Sim}, {Hirai}, {Podsiadlowski}, {Anderson}, {Ashall}, {Bate}, {Fender}, {Guti{\'e}rrez}, {Howell}, {Huber}, {Inserra}, {Leloudas}, {Monard}, {M{\"u}ller-Bravo}, {Shappee}, {Smith}, {Terreran}, {Tonry}, {Tucker}, {Young}, {Aamer}, {Chen}, {Ragosta}, {Galbany}, {Gromadzki}, {Harvey}, {Hoeflich}, {McCully}, {Newsome}, {Gonzalez}, {Pellegrino}, {Ramsden}, {P{\'e}rez-Torres}, {Ridley}, {Sheng}, \& {Weston}}]{Moore2023ApJ}
{Moore}, T., {Smartt}, S.~J., {Nicholl}, M., {et~al.} 2023, \bibinfo{title}{{SN 2022jli: A Type Ic Supernova with Periodic Modulation of Its Light Curve and an Unusually Long Rise},} \apjl, 956, L31, \dodoi{10.3847/2041-8213/acfc25}

\bibitem[{C.~D. {Murray} \& S.~F. {Dermott}(1999){Murray} \& {Dermott}}]{Murray1999ssd}
{Murray}, C.~D., \& {Dermott}, S.~F. 1999, {Solar System Dynamics} (Cambridge University Press), \dodoi{10.1017/CBO9781139174817}

\bibitem[{N. {Murray} {et~al.}(2011){Murray}, {M{\'e}nard}, \& {Thompson}}]{Murray2011ApJ}
{Murray}, N., {M{\'e}nard}, B., \& {Thompson}, T.~A. 2011, \bibinfo{title}{{Radiation Pressure from Massive Star Clusters as a Launching Mechanism for Super-galactic Winds},} \apj, 735, 66, \dodoi{10.1088/0004-637X/735/1/66}

\bibitem[{S. {Naoz}(2016){Naoz}}]{Naoz2016ARA&A}
{Naoz}, S. 2016, \bibinfo{title}{{The Eccentric Kozai-Lidov Effect and Its Applications},} \araa, 54, 441, \dodoi{10.1146/annurev-astro-081915-023315}

\bibitem[{E.~R. {Parkin} \& J.~M. {Pittard}(2008){Parkin} \& {Pittard}}]{Parkin2008MNRAS}
{Parkin}, E.~R., \& {Pittard}, J.~M. 2008, \bibinfo{title}{{A 3D dynamical model of the colliding winds in binary systems},} \mnras, 388, 1047, \dodoi{10.1111/j.1365-2966.2008.13511.x}

\bibitem[{N.~D. {Richardson} {et~al.}(2025){Richardson}, {Henson}, {Lieb}, {Kehl}, {Lau}, {Williams}, {Corcoran}, {Callingham}, {Chen{\'e}}, {Gull}, {Hamaguchi}, {Han}, {Hankins}, {Hill}, {Hoffman}, {Mackey}, {Moffat}, {Pope}, {Pradhan}, {Russell}, {Sander}, {St-Louis}, {Stevens}, {Tuthill}, {Weigelt}, \& {White}}]{Richardson2025ApJ}
{Richardson}, N.~D., {Henson}, M., {Lieb}, E.~P., {et~al.} 2025, \bibinfo{title}{{Carbon-rich Dust Injected into the Interstellar Medium by Galactic WC Binaries Survives for Hundreds of Years},} \apj, 987, 160, \dodoi{10.3847/1538-4357/addf30}

\bibitem[{L.~F. {Rodr{\'\i}guez} {et~al.}(2020){Rodr{\'\i}guez}, {Arthur}, {Montes}, {Carrasco-Gonz{\'a}lez}, \& {Toal{\'a}}}]{Rodriguez2020ApJ}
{Rodr{\'\i}guez}, L.~F., {Arthur}, J., {Montes}, G., {Carrasco-Gonz{\'a}lez}, C., \& {Toal{\'a}}, J.~A. 2020, \bibinfo{title}{{A Radio Pinwheel Emanating from WR 147},} \apjl, 900, L3, \dodoi{10.3847/2041-8213/abad9d}

\bibitem[{T. {Schirmer} {et~al.}(2022){Schirmer}, {Ysard}, {Habart}, {Jones}, {Abergel}, \& {Verstraete}}]{Schirmer2022A&A}
{Schirmer}, T., {Ysard}, N., {Habart}, E., {et~al.} 2022, \bibinfo{title}{{Nano-grain depletion in photon-dominated regions},} \aap, 666, A49, \dodoi{10.1051/0004-6361/202243635}

\bibitem[{A. {Soulain} {et~al.}(2023){Soulain}, {Lamberts}, {Millour}, {Tuthill}, \& {Lau}}]{Soulain2023MNRAS}
{Soulain}, A., {Lamberts}, A., {Millour}, F., {Tuthill}, P., \& {Lau}, R.~M. 2023, \bibinfo{title}{{Smoke on the wind: dust nucleation in the archetype colliding-wind pinwheel WR 104},} \mnras, 518, 3211, \dodoi{10.1093/mnras/stac2999}

\bibitem[{A. {Soulain} {et~al.}(2018){Soulain}, {Millour}, {Lopez}, {Matter}, {Lagadec}, {Carbillet}, {La Camera}, {Lamberts}, {Langlois}, {Milli}, {Avenhaus}, {Magnard}, {Roux}, {Moulin}, {Carle}, {Sevin}, {Martinez}, {Abe}, \& {Ramos}}]{Soulain2018A&A}
{Soulain}, A., {Millour}, F., {Lopez}, B., {et~al.} 2018, \bibinfo{title}{{SPHERE view of Wolf-Rayet 104. Direct detection of the Pinwheel and the link with the nearby star},} \aap, 618, A108, \dodoi{10.1051/0004-6361/201832817}

\bibitem[{J.~D. {Thomas} {et~al.}(2021){Thomas}, {Richardson}, {Eldridge}, {Schaefer}, {Monnier}, {Sana}, {Moffat}, {Williams}, {Corcoran}, {Stevens}, {Weigelt}, {Zainol}, {Anugu}, {Le Bouquin}, {ten Brummelaar}, {Campos}, {Couperus}, {Davies}, {Ennis}, {Eversberg}, {Garde}, {Gardner}, {Fl{\'o}}, {Kraus}, {Labdon}, {Lanthermann}, {Leadbeater}, {Lester}, {Maki}, {McBride}, {Ozuyar}, {Ribeiro}, {Setterholm}, {Stober}, {Wood}, \& {Zurm{\"u}hl}}]{Thomas2021MNRAS}
{Thomas}, J.~D., {Richardson}, N.~D., {Eldridge}, J.~J., {et~al.} 2021, \bibinfo{title}{{The orbit and stellar masses of the archetype colliding-wind binary WR 140},} \mnras, 504, 5221, \dodoi{10.1093/mnras/stab1181}

\bibitem[{S. {Toonen} {et~al.}(2016){Toonen}, {Hamers}, \& {Portegies Zwart}}]{Toonen2016ComAC}
{Toonen}, S., {Hamers}, A., \& {Portegies Zwart}, S. 2016, \bibinfo{title}{{The evolution of hierarchical triple star-systems},} Computational Astrophysics and Cosmology, 3, 6, \dodoi{10.1186/s40668-016-0019-0}

\bibitem[{P.~G. {Tuthill} {et~al.}(1999){Tuthill}, {Monnier}, \& {Danchi}}]{Tuthill1999Natur}
{Tuthill}, P.~G., {Monnier}, J.~D., \& {Danchi}, W.~C. 1999, \bibinfo{title}{{A dusty pinwheel nebula around the massive star WR104},} \nat, 398, 487, \dodoi{10.1038/19033}

\bibitem[{P.~G. {Tuthill} {et~al.}(2008){Tuthill}, {Monnier}, {Lawrance}, {Danchi}, {Owocki}, \& {Gayley}}]{Tuthill2008ApJ}
{Tuthill}, P.~G., {Monnier}, J.~D., {Lawrance}, N., {et~al.} 2008, \bibinfo{title}{{The Prototype Colliding-Wind Pinwheel WR 104},} \apj, 675, 698, \dodoi{10.1086/527286}

\bibitem[{V.~V. {Usov}(1991){Usov}}]{Usov1991MNRAS}
{Usov}, V.~V. 1991, \bibinfo{title}{{Stellar wind collision and dust formation in long-period, heavily interacting Wolf-Rayet binaries.},} \mnras, 252, 49, \dodoi{10.1093/mnras/252.1.49}

\bibitem[{A.~J. {\lowercase{V}an Marle} {et~al.}(2008){\lowercase{V}an Marle}, {Langer}, {Yoon}, \& {Garc{\'\i}a-Segura}}]{van-Marle2008A&A}
{\lowercase{V}an Marle}, A.~J., {Langer}, N., {Yoon}, S.~C., \& {Garc{\'\i}a-Segura}, G. 2008, \bibinfo{title}{{The circumstellar medium around a rapidly rotating, chemically homogeneously evolving, possible gamma-ray burst progenitor},} \aap, 478, 769, \dodoi{10.1051/0004-6361:20078802}

\bibitem[{P. Virtanen {et~al.}(2020)Virtanen, Gommers, Oliphant, Haberland, Reddy, Cournapeau, Burovski, Peterson, Weckesser, Bright, {van der Walt}, Brett, Wilson, Millman, Mayorov, Nelson, Jones, Kern, Larson, Carey, Polat, Feng, Moore, {VanderPlas}, Laxalde, Perktold, Cimrman, Henriksen, Quintero, Harris, Archibald, Ribeiro, Pedregosa, {van Mulbregt}, \& {SciPy 1.0 Contributors}}]{SciPy2020NaturMth}
Virtanen, P., Gommers, R., Oliphant, T.~E., {et~al.} 2020, \bibinfo{title}{{{SciPy} 1.0: Fundamental Algorithms for Scientific Computing in Python},} Nature Methods, 17, 261, \dodoi{10.1038/s41592-019-0686-2}

\bibitem[{D.~J. {Wallace} {et~al.}(2002){Wallace}, {Moffat}, \& {Shara}}]{Wallace2002ASPC}
{Wallace}, D.~J., {Moffat}, A. F.~J., \& {Shara}, M.~M. 2002, \bibinfo{title}{{Hubble Space Telescope Detection of Binary Companions Around Three WC9 Stars: WR 98a, WR 104, and WR 112},} in Astronomical Society of the Pacific Conference Series, Vol. 260, Interacting Winds from Massive Stars, ed. A.~F.~J. {Moffat} \& N.~{St-Louis}, 407

\bibitem[{R.~M.~T. White(2025)White}]{White2025Zenodo}
White, R. M.~T. 2025, xenomorph: the fastest way to model the colliding winds of massive binary stars, Zenodo, \dodoi{10.5281/zenodo.15875502}

\bibitem[{R.~M.~T. {White} \& P. {Tuthill}(2024){White} \& {Tuthill}}]{White2024arXiv}
{White}, R. M.~T., \& {Tuthill}, P. 2024, \bibinfo{title}{{Wolf-Rayet Colliding Wind Binaries},} arXiv e-prints, arXiv:2412.12534, \dodoi{10.48550/arXiv.2412.12534}

\bibitem[{P.~M. {Williams} {et~al.}(2009{\natexlab{a}}){Williams}, {Rauw}, \& {van der Hucht}}]{Williams2009MNRASb}
{Williams}, P.~M., {Rauw}, G., \& {van der Hucht}, K.~A. 2009{\natexlab{a}}, \bibinfo{title}{{Dust formation by the colliding wind WC5+O9 binary WR19 at periastron passage},} \mnras, 395, 2221, \dodoi{10.1111/j.1365-2966.2009.14681.x}

\bibitem[{P.~M. {Williams} {et~al.}(2012){Williams}, {van der Hucht}, {van Wyk}, {Marang}, {Whitelock}, {Bouchet}, \& {Setia Gunawan}}]{Williams2012MNRAS}
{Williams}, P.~M., {van der Hucht}, K.~A., {van Wyk}, F., {et~al.} 2012, \bibinfo{title}{{Recurrent dust formation by WR 48a on a 30-year time-scale},} \mnras, 420, 2526, \dodoi{10.1111/j.1365-2966.2011.20218.x}

\bibitem[{P.~M. {Williams} {et~al.}(2009{\natexlab{b}}){Williams}, {Marchenko}, {Marston}, {Moffat}, {Varricatt}, {Dougherty}, {Kidger}, {Morbidelli}, \& {Tapia}}]{Williams2009MNRAS}
{Williams}, P.~M., {Marchenko}, S.~V., {Marston}, A.~P., {et~al.} 2009{\natexlab{b}}, \bibinfo{title}{{Orbitally modulated dust formation by the WC7+O5 colliding-wind binary WR140},} \mnras, 395, 1749, \dodoi{10.1111/j.1365-2966.2009.14664.x}

\bibitem[{S.~A. {Zhekov} \& B.~V. {Petrov}(2025){Zhekov} \& {Petrov}}]{Zhekov2025A&A}
{Zhekov}, S.~A., \& {Petrov}, B.~V. 2025, \bibinfo{title}{{Chandra high-resolution spectra of Apep Plume: Exploring the colliding-wind picture in X-rays},} \aap, 693, A266, \dodoi{10.1051/0004-6361/202452599}

\end{thebibliography}
